\documentclass[journal,comsoc]{IEEEtran}
\usepackage{times}

\usepackage{physics}
\usepackage{hyperref}
\usepackage{graphicx}
\usepackage[font=footnotesize]{caption}
\usepackage{amsmath}
\usepackage{amsfonts}
\usepackage{amssymb}
\usepackage{amsthm}
\usepackage{tabularx}
\usepackage{mathrsfs} 
\usepackage{times}
\usepackage{cite}

\usepackage{subfig}

\usepackage{soul}

\usepackage{stfloats}
\usepackage{float}
\usepackage{url}
\usepackage{hyperref}
\newtheorem{theorem}{Theorem}
\newtheorem{definition}{Definition}
\newtheorem{lemma}{Lemma}
\newtheorem{corollary}{Corollary}

\newtheorem{proposition}{Proposition}
\newtheorem{approximation}{Approximation}

\usepackage[nolist,nohyperlinks]{acronym}
\usepackage[numbers,sort&compress]{natbib}

\usepackage{bbm}

\usepackage{array}
\usepackage{booktabs}
\usepackage{colortbl}
\usepackage{mdwmath}
\usepackage{mdwtab}
\usepackage{eqparbox}
\usepackage{cellspace}
\usepackage{makecell}

\DeclareGraphicsExtensions{.png,.eps,.ps,.pdf}

\usepackage{tikz}
\usetikzlibrary{arrows,    
                shapes, 
                positioning, 
                fit,    
                backgrounds, 
                mindmap,
                petri,
                intersections, 
                external %
                }
\definecolor{emerald}{RGB}{69,155,61}
\definecolor{gold}{RGB}{244,216,51}
\definecolor{pink}{RGB}{235,44,206}
\usepackage{verbatim}

\begin{acronym}
\acro{SNR}{signal-to-noise ratio}
\acrodefplural{SNR}[SNRs]{signal-to-noise ratios}
\acro{RV}{random variable}
\acrodefplural{RV}[RVs]{random variables}
\acro{LOS}{line-of-sight}
\acrodefplural{LoS}{Line-of-sight}
\acro{EH}{Energy Harvesting}
\acro{PDF}{probability density function}
\acro{CDF}{cumulative distribution function}
\acrodefplural{CDF}[CDFs]{cumulative distribution functions}
\acro{IoT}{Internet of Things}
\acro{PB}{Power Beacon}
\acrodefplural{PB}[PBs]{Power Beacons}
\acro{WPC}{Wireless Powered Communications}
\acro{WPT}{Wireless Power Transmission}
\acro{RFID}{Radio Frequency IDentification}
\acro{MC}{Monte Carlo}
\acro{PLS}{Physical Layer Security}
\acro{SWIPT}{Simultaneous Wireless and Information Power Transfer}
\end{acronym}

\tikzstyle{int}=[draw, fill=cyan!20, minimum size=2em]
\tikzstyle{int_blue}=[draw, fill=black!20, minimum size=2em]
\tikzstyle{int_green}=[draw, fill=green!20, minimum size=2em]
\tikzstyle{int_red}=[draw, fill=red!20, minimum size=2em]
\tikzstyle{init} = [pin edge={to-,thin,black}]

\begin{document}
\title{Joint Distribution of Distance and Angles in Finite Wireless Networks}
\author{Francisco J. Mart\'in-Vega, Gerardo G\'omez, David Morales-Jim\'enez, F. Javier L\'opez-Mart\'inez and Mari~Carmen~Aguayo-Torres
\thanks{Manuscript received April xx, 2022; revised XXX. %
This work has been funded by the European Fund for Regional Development (FEDER), AEI (Spain), Junta de Andaluc\'ia and the University of M\'alaga through grants RYC2021-034620-I, and RYC2020-030536-I, UMA20-FEDERJA-002, TIC102 research group, and by MCIN/AEI/10.13039/501100011033 through grant PID2020-118139RB-I00.}
\thanks{The authors are with the Communications and Signal Processing Lab, Telecommunication Research Institute (TELMA), Universidad de M\'alaga, E.T.S. Ingenier\'ia de Telecomunicaci\'on, Bulevar Louis Pasteur 35, 29010 M\'alaga (Spain). Dr. Morales-Jim\'enez and Dr. L\'opez-Mart\'inez are also with the Department of Signal Theory, Networking and Communications, University of Granada, 18071 Granada (Spain). (e-mail: fjmvega@ic.uma.es)}
}

\markboth{IEEE TRANSACTIONS ON VEHICULAR TECHNOLOGY, VOL. XXX, NO. XXX, MAY 2023}%
{Shell \MakeLowercase{\textit{et al.}}: Bare Demo of IEEEtran.cls for IEEE Communications Society Journals}

\maketitle
\begin{abstract}
Directional beamforming will play a paramount role in 5G and beyond networks to combat the higher path losses incurred at millimeter wave bands.  Appropriate modeling and analysis of the angles and distances between transmitters and receivers in these networks are thus essential to understand performance and limiting factors. Most existing literature considers either infinite and uniform networks, where nodes are drawn according to a Poisson point process, or finite networks with the reference receiver placed at the origin of a disk. Under either of these assumptions, the distance and azimuth angle between transmitter and receiver are independent, and the angle follows a uniform distribution between $0$ and $2\pi$. Here, we consider a more realistic case of finite networks where the reference node is placed at any arbitrary location. We obtain the joint distribution between the distance and azimuth angle and demonstrate that these random variables do exhibit certain correlation, which depends on the shape of the region and the location of the reference node. To conduct the analysis, we present a general mathematical framework {that} is specialized to exemplify the case of a rectangular region. We also derive the statistics for the 3D case where, considering antenna heights, the joint distribution of distance, azimuth, and zenith angles is obtained. Finally, we describe some immediate applications of the present work, including the design of analog codebooks, wireless routing algorithms, and the analysis of directional beamforming{, which is illustrated by analyzing the coverage probability of an indoor scenario considering misaligned beams.} %
\end{abstract}


\section{Introduction}
\label{sec:Intro}

\subsection{Motivation and Scope}
\IEEEPARstart{T}{he} need {for} greater bandwidths to accommodate the ever{-}increasing demand of data rates {has} led to the use of higher frequency bands, e.g., millimeter wave (mmW) bands. {The} key to compensate {for} the higher path loss experienced at these bands is directional beamforming, which uses a massive number of antenna elements that can be conveniently packed due to the smaller {wavelengths}. The enabling role of directional beamforming in the realization of 5G and beyond networks is unquestionable, and appropriate modeling and analysis {are} needed to identify performance trends, trade-offs, and limiting factors \cite{ElSawy17, DiRenzo21, Martin18_Mag}. {For that reason}, {some} works {address previous problems by} analyzing directional beamforming on a {broad} set of scenarios such as 5G cellular networks 
\cite{Shi21, Kalamkar21, Rebato19}, 
vehicular networks \cite{Singh21, Ammar20, Yuanwei19}, 
device-to-device (D2D) communications \cite{Kusaladharma19, Haenggi18, Shuanshuan18}, 
unmanned aerial vehicles (UAVs) based networks \cite{Maeng21, Azari20, Enayati19}, and 
wireless communications empowered with reconfigurable intelligent surfaces (RISs) \cite{Lyu21, Alouini21, Nemati21}. 

In these works, assuming a random location of the transmitters/receivers, the analyses typically require the joint statistical distribution of the polar coordinates (distance and azimuth angle) in 2D scenarios; and that of the spherical coordinates, which also include the zenith (elevation) angle in 3D scenarios. The latter case is considered when the height of the nodes is relevant (e.g., UAV scenarios). In 2D scenarios{,} it is assumed that the distance, $R$, and azimuth angle, $\Theta$, between the transmitter and receiver are independent random variables (RVs), where $\Theta$ follows a uniform distribution between $0$ and $2\pi$. Those works investigating 3D scenarios make the same assumptions for the distance and azimuth angle, but they consider the zenith angle, $\Psi$, correlated with the distance. Such correlation comes from the fact that $\Psi$ can be expressed in simple terms as a trigonometric function of the distance and the height of the antennas.  


While these assumptions 
are valid for infinite networks with uniform node distributions, e.g., governed by a uniform Poisson Point Process (PPP), they do not hold for finite networks. Indeed, as shown in this work, the azimuth angle and distance between a randomly placed node and the reference node are correlated {in finite networks}. Modeling this correlation is very challenging as it depends on both the shape of the region (network) where the nodes are located and on the position of the reference node. 
The only exception is the case 
where the reference node is at the center of a disk; in such case, the azimuth angles are independent of the distance{,} and follow a uniform distribution within $[0, 2\pi]$.

\subsection{Related Work}

Considering finite networks is crucial to investigate the system-level performance of directional beamforming, which typically aims at boosting the throughput in limited regions with an increased demand for high data rates (hot spots). Nevertheless, the analysis of finite networks, where node locations are modeled by a binomial point process (BPP), is substantially more complex than that of infinite networks modeled by a PPP. In particular: i) in finite networks, the distribution of the signal to interference plus noise ratio (SINR) depends on the position of the reference node and,  therefore, the performance is location dependent; ii) the distances of all nodes in the network towards a reference node are correlated; iii) unlike for PPP-based infinite networks, the selection of a randomly placed node in finite networks changes the distribution of the underlying point process \cite{Afshang17}.

Despite its relevance, the analysis of directional beamforming in finite networks is scarce. 
{Most of the existing works avoid the need to compute the joint distribution of distance and angle, either because they consider a disk with the receiver placed at the center, or because they consider perfect beam-alignment for the desired link with sectored antenna patterns.} 
{For instance,} \cite{DaSilva20} {considers} that the transmitting nodes are randomly placed within a disk to analyze the 2D indoor mmW {case}. The probe receiver is placed at a given distance from the center of the disk and perfect beam alignment is assumed between the probe transmitter and receiver. However, the steering directions of the interfering beams are {considered} uniform and independent, thus ignoring the said distance-angle correlation. 
The uplink of an indoor wireless system operating at the terahertz band is analyzed in \cite{Kokkoniemi20}, considering a 3D rectangular region. The probe receiver is arbitrarily placed{,} whereas the interfering transmitters are randomly placed within the 3D region. Nevertheless, all transmit and receive beams are assumed to be perfectly aligned. Again, this avoids the need to compute the joint distribution and neglects the said correlation.

{In \mbox{\cite{Abarghouyi18a}}, a framework for analyzing single and multi-cluster wireless network is presented for the case of omnidirectional  antennas. For single clustered networks, active transmitters are assumed to be placed within a disk of radius $D$ following a finite homogeneous Poisson point process (FHPP) whereas the reference receiver is placed at a distance $d$ from the disk center. For multi-cluster networks, a Matern cluster process is assumed. Two association strategies are considered for each type of network to derive the coverage probability. }

{An interesting framework is presented in} \cite{Azimi-Abarghouyi19}, where 2D finite mmW systems with sectored antenna patterns are investigated. {This works assumes a maximum average received power association criteria and a considers a distance-dependent blockage probability model that categorizes the links as LOS and NLOS.}
{Again, the transmitters and receivers are assumed to be placed within a disk, and the reference receiver is placed at any arbitrary location within the disk, thus taking the correlation between angle and distance into account. However, perfect beam alignment is assumed for the desired link, avoiding the need to determine the joint distribution of distance and angle. The analysis considers the misalignment between interfering transmitters and the reference receiver. However, thanks to the assumed sectored pattern, only the probability mass function of the beamforming gain is needed, thus circumventing the need to compute the said joint distribution.} 


Moreover, existing results on the distance distributions in finite regions do not account for the angle distribution \cite{Srinivasa10, Khalid13, Fan07}

\subsection{Main Contributions}
To our knowledge, the azimuth angle distribution has not been derived yet for finite networks, despite its relevance to model and analyze directional beamforming {accurately}. This {issue} motivated us to investigate the correlation between the distance and azimuth angle. The main contributions of this work are:
\begin{enumerate}
    \item We present a mathematical framework to obtain the joint distribution of the distance and azimuth angle in 2D arbitrarily shaped networks. We consider the angle and distance between an arbitrary (fixed) point and a uniformly distributed random point.
    %
    \item We particularize the proposed framework to the case of a rectangular region, since this kind of region matches many practical scenarios of interest. 
    We derive the joint cumulative density function (CDF) and joint probability density function (PDF) of the distance and angle, and compute the marginals for the azimuth angle.
    \item We extend the obtained results to the case of 3D networks, {considering} both the height of the reference node and that of the random node. 
    %
    %
    \item Finally, we discuss some of the applications of the present work, including the analysis of directional beamforming, the design of the beam patterns of analog codebooks{,} and the design of wireless routing algorithms. {We illustrate the usefulness of the expressions by analyzing the coverage probability in an indoor scenario and determining the optimal AP location and antenna's bearing angle}. 
    %
\end{enumerate}

The derivation of the joint distance and azimuth angle distribution is significantly more challenging than the marginal distance distribution considered in the existing literature. For the derivation of the marginal distance distribution, say, in a regular polygon, one only needs to compute the intersection between the polygon (region of interest) and a disk centered at the reference node. To derive the joint distribution, however, we need to compute the intersection of $3$ regions: the region of interest, a disk of radius $r${,} and a given sector. Besides, extending the results to the 3D case requires the inclusion of an additional angle, the zenith (elevation) angle, which complicates the analysis. Moreover, it should be noted that the analysis of a rectangular region is substantially more complex than the case of a regular L-polygon or a disk, {as} rotational symmetry properties no longer hold.


\subsection{Notation and Paper Organization}
The following notation is used through{out} the text. $\mathbb{Z}$ and $\mathbb{R}$ {stand} for the set of integers and real numbers{,} respectively, $\mathbb{R}^+$ represents the non-negative real numbers{,} 
and $\mathbb{R}^d$ stands for the $d$-dimensional Euclidean space. Other sets are represented with fraktur font, e.g., $\mathfrak{R} \subseteq \mathbb{R}^d$, with $d > 0$, whereas boolean expressions are written with calligraphic font, e.g., $\mathcal{C} = \{ x \leq a\}$. The Lebesgue measure of the set $\mathfrak{R}$ is denoted by
$\lvert \mathfrak{R} \lvert$, which represents the area or the volume for $d=2$ or $d=3$, respectively. The intersection and union of sets are represented with symbols $\cap$ and $\cup$, respectively, whereas $\wedge$ and $\vee$ stand for the \texttt{and} and \texttt{or} logical operations. The empty set is denoted by $\emptyset$, whereas $\bar{\mathfrak{R}}$ represents the complement of the set $\mathfrak{R}$. In addition, the upper bar represents {the} negation of a logical expression, i.e., $\bar{\mathcal{C}}$ is \textit{true} if $\mathcal{C}$ is \textit{false}. 
If $X$ is a RV, then $F_X(x) = \Pr(X \leq x)$ represents its CDF where $\Pr(\cdot)$ denotes the probability measure.


\section{Mathematical Framework}
\label{sec:math_framework}

\subsection{Mathematical Preliminaries and Definitions}

\begin{figure}[t]
    \centering
    \includegraphics[width=0.55\columnwidth]{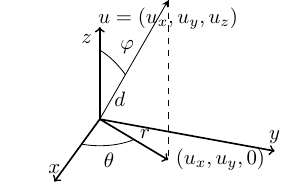}
    \caption{Coordinate system including: 1) spherical coordinates for distance, $d$, azimuth angle, $\theta$, and zenith angle, $\varphi$, related to a 3D point placed at $u=(u_x, u_y, u_z)$; and 2) polar coordinates for distance, $r$, and azimuth angle, $\theta$, related to the projection of the 3D point in the $xy$ plane.}
    \label{fig:coordinateSystem}
\end{figure}


We consider points that belong to either the 2D or 3D Euclidean space. They are represented using Cartesian coordinates, {in} either polar (2D) or spherical (3D) coordinates. Fig. \ref{fig:coordinateSystem} illustrates the spherical and polar coordinate systems, formalized in the {following} two definitions. Note that 2D points can be seen as projections of 3D points in the $xy$ plane.

\begin{definition}[Spherical coordinate system]
    \label{def:spherical_coord}
    An arbitrary point $u \in \mathbb{R}^3$ expressed as $(u_x,u_y,u_z)$ in Cartesian coordinates can be written as $(\mathscr{d}, \theta, \varphi)$, being $\mathscr{d}$ the distance, $\theta$ the azimuth angle, and $\varphi$ the zenith angle. 
    The following relations hold: 
    \begin{align}
    \label{eq:spherical_coord}
    u_x = \mathscr{d} \cos (\theta) \sin (\varphi); \; u_y = \mathscr{d} \sin (\theta) \sin (\varphi); \; u_z = \mathscr{d} \cos (\varphi).
    \end{align}
    %
    %
\end{definition}

\begin{definition}[Polar coordinate system]
    \label{def:polar_coord}
    An arbitrary point $u \in \mathbb{R}^2$ expressed as $(u_x,u_y)$ in Cartesian coordinates can be written as $(r, \theta)$, being $r$ the distance and $\theta$ the azimuth angle. 
    The following relations hold: 
    \begin{align}
    \label{eq:polar_coord}
    x = r \cos (\theta); \; y = r \sin (\theta);
    \end{align}
\end{definition}

We present some mathematical functions used {to analyze} joint angle and distance distributions. 

\begin{definition}[Dirac Delta function]
    \label{def:DiracDelta}
    {The Dirac Delta function, $\delta(x)$, is described as a generalized function that fulfills these two conditions:}
    \textcolor{black}{\begin{align}
        \label{eq:delta_cond1}
        &\delta (x) = \lim_{\epsilon \to 0^+} g(x, \epsilon)  = 
        \left\{
        \begin{array}{ll}
            \infty & \mathrm{if} \quad x = 0 \\
            0 & \mathrm{otherwise} 
        \end{array} 
        \right. \\
        &\int_{-\infty}^{\infty} \delta (x) \dd{x} = \lim_{\epsilon \to 0^+} \int_{-\infty}^{\infty} g(x, \epsilon) \dd{x} = 1
    \end{align}}
    \noindent {where $g(x, \epsilon)$ can be any function satisfying the above equations.}
\end{definition}

\begin{definition}[Indicator function]
    \label{def:indicator}
    The indicator function evaluated over a set, $\mathfrak{C} \subset \mathbb{R}^d$, with $d>0$, is defined as 
    \begin{equation}
        \mathbbm{1}_{\mathfrak{C}}(x) = \int_{\mathfrak{C}} \delta (x - u) \dd{u} =
        \left\{
        \begin{array}{ll}
            1 & \mathrm{if} \quad x \in \mathfrak{C} \\
            0 & \mathrm{otherwise} 
        \end{array} 
        \right. 
    \end{equation}
    \noindent where $\delta(x)$ is the Dirac delta function in $\mathbb{R}^d$.
\end{definition}

The expression $x \in \mathfrak{C}$ in the above definition can also be viewed as a boolean expression representing an event\footnote{The indicator function of a subset (or event) maps elements of the subset (i.e., the event that $x$ falls in the subset) to one and zero otherwise.}, $\mathcal{C}$, that is \textit{true} if $x$ belongs to the set $\mathfrak{C}$ and \textit{false} otherwise. Thus, the indicator function can alternatively be written as $\mathbbm{1}_{\mathfrak{C}}(x) = \mathbbm{1}\left(\mathcal{C}\right)$ \cite{Haenggi12}, where $\mathcal{C} = \{x \in \mathfrak{C}\}$. 

\begin{definition}[Properties of the indicator function] \label{properties_if}
The logical \texttt{and} ($\wedge$) and \texttt{or} ($\vee$) operations on two boolean expressions, $\mathcal{C}_1$ and $\mathcal{C}_2$, satisfy the following relations:
\begin{align}
    \mathbbm{1} (\mathcal{C}_1 \wedge \mathcal{C}_2) &= \mathbbm{1} (\mathcal{C}_1) \mathbbm{1} (\mathcal{C}_2), \nonumber \\
    \mathbbm{1} (\mathcal{C}_1 \vee \mathcal{C}_2) &= \mathbbm{1} (\mathcal{C}_1) + \mathbbm{1} (\mathcal{C}_2) - \mathbbm{1} (\mathcal{C}_1 \wedge \mathcal{C}_2).
\end{align}
Equivalently, the union and intersection of two sets, $\mathfrak{C}_1$ and $\mathfrak{C}_2$, lead to the following equalities:
\begin{align}
    \mathbbm{1}_{\mathfrak{C}_1 \cap \mathfrak{C}_2}(x) &= \mathbbm{1}_{\mathfrak{C}_1}(x) \mathbbm{1}_{\mathfrak{C}_2}(x), \nonumber \\
    \mathbbm{1}_{\mathfrak{C}_1 \cup \mathfrak{C}_2}(x) &= \mathbbm{1}_{\mathfrak{C}_1}(x) + \mathbbm{1}_{\mathfrak{C}_2}(x) - \mathbbm{1}_{\mathfrak{C}_1 \cap \mathfrak{C}_2}(x).
\end{align}
\end{definition}

The {following} two definitions allow {writing} the main mathematical expressions obtained in this work in compact form. 

\begin{definition}[Positive part operator]
    \label{def:+ operator}
    For any real number (or real-valued function) $x$, the positive part of $x$ is defined as
    %
    \begin{equation}
        (x)^+ = \max\{0,x\} =
        \left\{
        \begin{array}{ll}
            x & \mathrm{if} \quad x > 0, \\
            0 & \mathrm{otherwise}.
        \end{array} 
        \right. 
    \end{equation}
    
\end{definition}

\begin{definition}[$\mathbbm{F}$ operator]
    \label{def:F operator}
    For a function $f$ on the real domain, i.e., $f: \mathbb{R} \mapsto \mathbb{R}$, the operator $\mathbbm{F} (f; a; b)$ is defined as
    \begin{equation}
        \label{eq:F}
        \mathbbm{F} (f; a; b) =
        \left\{
        \begin{array}{ll}
            f(b) - f(a) & \mathrm{if} \quad b \geq a, \\
            0 & \mathrm{otherwise}.
        \end{array} 
        \right. 
    \end{equation}
    %
\end{definition}
This definition allows us to write the result of definite integrals in compact form, as illustrated next. 

\begin{proposition}
    \label{prop:Definite_int}
    The definite integral $\int_{b_1(u)}^{b_2(u)} \mathbbm{1}_\mathfrak{A}(t) g(t) \dd{t}$, with $g(t) = \dv{f(t)}{t}$ and $\mathfrak{A} = [a_1(v), a_2(v)]$ admits
    \begin{align}
        \int_{b_1(u)}^{b_2(u)} &\mathbbm{1}_\mathfrak{A}(t) g(t) \dd{t} = \nonumber \\ 
            & \mathbbm{F}\left(f; \max(a_1(v), b_1(u)), \min(a_2(v), b_2(u)) \right).
    \end{align}
    \begin{proof}
        The proof comes after (i) applying the indicator function to the integration limits; (ii) considering that if $a_1(v) > a_2(v)$, then $\mathfrak{A}$ reduces to an empty set and the result of the integral is $0$; and (iii) substituting the $\mathbbm{F}$ operator in the resulting expression. 
    \end{proof}
\end{proposition}

\begin{corollary}
    \label{cor:Definite_int}
    If $f(t)$ is a non-decreasing function, then the definite integral in \textbf{Proposition \ref{prop:Definite_int}} can be further expressed as
    \begin{align}
        \int_{b_1(u)}^{b_2(u)} &\mathbbm{1}_\mathfrak{A}(t) g(t) \dd{t} = \nonumber \\ 
            &\Big(f(\min(a_2(v), b_2(u))) -  f(\max(a_1(v), b_1(u)))\Big)^+.
    \end{align}
\end{corollary}

\subsection{Joint distribution in arbitrarily shaped networks}
\label{sec:ArbitraryShaped}

In this section, we present {a general} procedure to derive the joint distribution of the distance, $R$, and azimuth angle, $\Theta$, between a random point in the arbitrary region of interest $\mathfrak{R}(o) \subset \mathbb{R}^2$, and the reference point $u=(u_x, u_y) \in \mathbb{R}^2$. The region is modeled as a set whose center of mass is located at the origin $o=(0,0)$, and random points are uniformly distributed within the region. 

\begin{figure}[t]
    \centering
    \includegraphics[width=0.65\columnwidth]{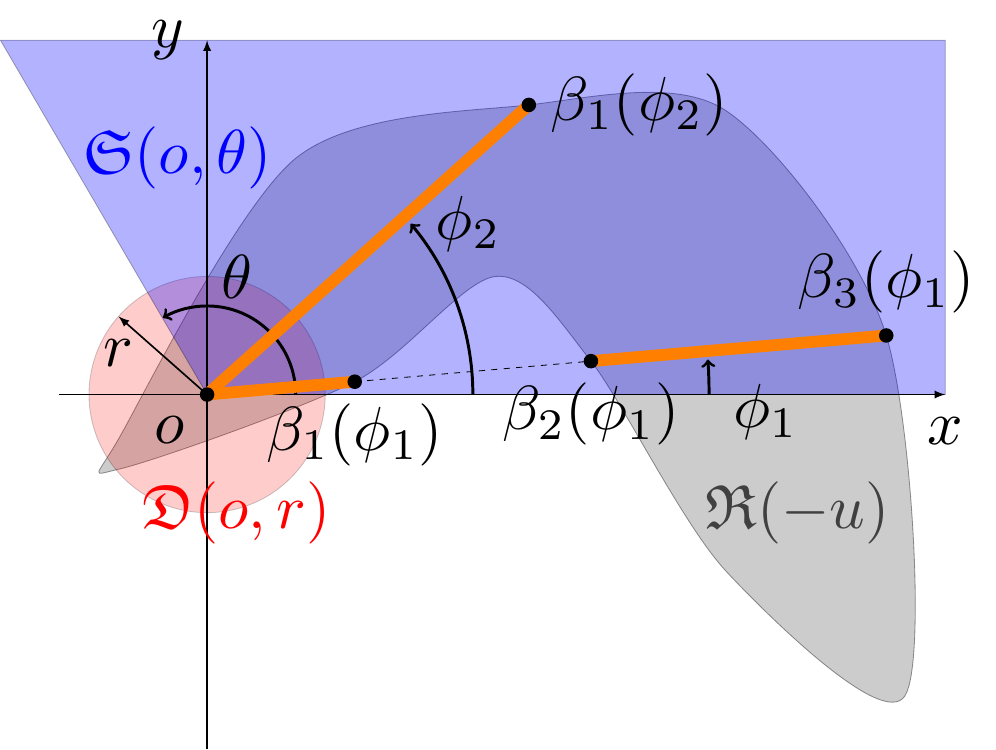}
    \caption{Sketch of the relevant sets involved in the derivation of the joint distribution of distance, $R$, and azimuth angle, $\Theta$, for points uniformly placed in an arbitrary set (the region of interest), $\mathfrak{R}(-u)$. The disk of radius $r$ centered at the origin $o=(0,0)$ (region in red) is denoted by $\mathfrak{D}(o, r)$, while $\mathfrak{S}(o, \theta)$ represents a sector with angle span $[0,\theta]$ (region in black). The set $\mathfrak{B}(\phi)$ is drawn in orange for the azimuth angles $\phi_1$ and $\phi_2$.}
    \label{fig:arbitraryShaped}
\end{figure}

The joint CDF of the distance and azimuth angle can be written as

\begin{align}
    \label{eq:F_R_Theta}
    F_{R, \Theta} (r, \theta) &= 
        \frac{\lvert \mathfrak{Z}(u,r,\theta) \lvert}{\lvert \mathfrak{R}(o)\lvert} 
    = \frac{\lvert \mathfrak{D}(u,r) \cap \mathfrak{R}(o) \cap \mathfrak{S}(u,\theta) \lvert}{\lvert \mathfrak{R}(o)\lvert} \nonumber \\
    &\overset{\mathrm{(a)}}{=} \frac{\lvert \mathfrak{D}(o,r) \cap \mathfrak{R}(-u) \cap \mathfrak{S}(o,\theta) \lvert}{\lvert \mathfrak{R}(-u)\lvert}.
\end{align}

\noindent where $\mathfrak{D}(u,r)$ is a disk of radius $r$ centered at $u$ and $\mathfrak{S}(u, \theta)$ is the sector spanning $[0,\theta]$ with origin at $u$. The set $\mathfrak{Z}(u,r,\theta)$ represents the intersection of both regions with $\mathfrak{R}(o)$ the region of interest, and (a) follows from the fact that the Lebesgue measure is invariant to translations. Fig. \ref{fig:arbitraryShaped} illustrates an example of those sets. The disk $\mathfrak{D}(o,r)$, sector $\mathfrak{S}(o,\theta)$, and the translated region of interest (centered at $-u$) $\mathfrak{R}(-u)$ are drawn in red, black and gray color, respectively. 

The main challenge to derive the joint CDF is the computation of the area of $\mathfrak{Z}(u,r,\theta)$, which can be expressed as

\begin{align}
    \lvert \mathfrak{Z}(u,r,\theta) \lvert &= \int_{\phi=0}^{2\pi} \int_{\rho=0}^{\infty} \rho 
        \mathbbm{1}_{\mathfrak{D}(o,r) \cap \mathfrak{R}(-u) \cap \mathfrak{S}(o,\theta)} \left( \rho, \theta \right) \dd{\rho} \dd{\phi} \nonumber \\
    &\overset{\mathrm{(a)}}{=} \int_{\phi=0}^{\theta} \int_{\rho=0}^{r} \rho 
    \mathbbm{1}_{\mathfrak{R}(-u)} \left( \rho, \theta \right) \dd{\rho} \dd{\phi} \nonumber \\
    &\overset{\mathrm{(b)}}{=} \int_{\phi=0}^{\theta} \int_{\rho=0}^{r} \rho 
\mathbbm{1}_{\mathfrak{B}(\phi)} \left( \rho \right) \dd{\rho} \dd{\phi},
\end{align}

\noindent where (a) comes after applying \textbf{Definition \ref{properties_if}} to the indicator function, and translating the resulting indicator functions for the disk and the sector into corresponding changes on the integration limits over $\rho$ and $\phi$, respectively; while (b) comes after the fact that $\mathfrak{R}(-u) = \{ \rho\in \mathbb{R}^+, \phi \in [0, 2\pi] \lvert \rho \in \mathfrak{B}(\phi) \}$.

The set $\mathfrak{B}(\phi)  \subset \mathbb{R}^+$ represents, for a given azimuth angle $\phi$, the range of values of $\rho$ that belong to the set $\mathfrak{R}(-u)$. Note that, if the arbitrary region is composed of concave sets and/or the union of disjoint sets, $\mathfrak{B}(\phi)$ will be the union of disjoint intervals, where the number of intervals and their limits depend on the azimuth angle, $\phi$. 
This is illustrated in Fig. \ref{fig:arbitraryShaped}. In the shown example, for $\phi_1$, we have $\mathfrak{B}(\phi_1) = (0, \beta_1(\phi_1)] \cap [\beta_2(\phi_1), \beta_3(\phi_1)]$; whereas, for $\phi_2$, $\mathfrak{B}(\phi_2) = (0, \beta_1(\phi_2)]$. 

If $\mathfrak{R}(-u)$ is a convex region, we can always write $\mathfrak{B}(\phi) = [0, \beta(\phi)]$, since the line segment between any two points in a convex set will always lie within the set \cite{Boyd06}. In such case,

\begin{align}
    \lvert \mathfrak{Z}(u,r,\theta) \lvert  
        &\overset{\mathrm{(a)}}{=} \int_{\phi=0}^{\theta} \int_{\rho=0}^{\min(r, \beta(\phi))} \rho  \dd{\rho} \dd{\phi} \nonumber \\
    &\overset{\mathrm{(b)}}{=} \frac{1}{2} \int_{\phi=0}^{\theta} r^2 \mathbbm{1} \big( \underbrace{ r \leq \beta(\phi)}_{\mathcal{B}(r,\phi)} \big) \dd{\phi} \nonumber \\
    &+ \frac{1}{2} \int_{\phi=0}^{\theta} \beta^2 (\phi)  \mathbbm{1} \big( \underbrace{ r > \beta(\phi)}_{\bar{\mathcal{B}}(r,\phi)} \big) \dd{\phi},
    \label{eq:int_convex}
\end{align}

\noindent where $\mathcal{B}(r,\phi)$ stands for the boolean expression (or event) \mbox{$r \leq \beta(\phi)$} and $\bar{\mathcal{B}}(r,\phi)$ represents its complement, i.e., \mbox{$r > \beta(\phi)$}. Note that \eqref{eq:int_convex} is significantly more tractable than the initial problem to compute the overlap area 
$\lvert \mathfrak{Z}(u,r,\theta) \lvert$. 


{The computation of \mbox{\eqref{eq:int_convex}} requires to express the limits of integration in a tractable form. This can be done in a systematic way by the following step-by-step procedure:}
\begin{itemize}
    \item \textbf{Step 1}:  Derive $\beta(\phi)$ to write the convex set
    $\mathfrak{R}(-u) =  \{ \rho\in \mathbb{R}^+, \phi \in [0, 2\pi] \lvert \mathcal{B}(\rho,\phi) \}$ with $\mathcal{B}(\rho,\phi) = \{\rho \leq \beta(\phi)\}$.
    \item \textbf{Step 2}: Solve the inequalities defining the events $\mathcal{B}(r,\phi)$ and $\bar{\mathcal{B}}(r,\phi)$, i.e., $r \leq \beta(\phi)$ and $r > \beta(\phi)$ to derive the disjoint sets of azimuth angles which satisfy either of the two events. These sets are written as 
    $\mathfrak{T_{\mathfrak{B}}}(r) = \{\phi \in [0, 2\pi] \lvert \mathcal{B}(r,\phi) = \mathtt{1} \}$ and $\mathfrak{T_{\bar{\mathfrak{B}}}}(r) = \{\phi \in [0, 2\pi] \lvert \bar{\mathcal{B}}(r,\phi) = \mathtt{1} \}$. 
    \item \textbf{Step 3}: Use $\mathbbm{1}(\mathcal{B}(r,\phi)) = \mathbbm{1}_{\mathfrak{T_{\mathfrak{B}}}(r)}(\phi)$ and $\mathbbm{1}(\bar{\mathcal{B}}(r,\phi)) = \mathbbm{1}_{\mathfrak{T_{\bar{\mathfrak{B}}}}(r)}(\phi)$ in \eqref{eq:int_convex} and modify the integration limits according to $\mathfrak{T_{\mathfrak{B}}}(r)$ and $\mathfrak{T_{\bar{\mathfrak{B}}}}(r)$ to solve the integrals. 
\end{itemize}

{Note that this procedure is general and applicable to any arbitrarily shaped convex region.}

In the next section{,} we follow the above procedure to derive the joint CDF of distance and angle in the relevant case of a rectangle. 



\section{Rectangular Regions}
\label{sec:Analysis}
\subsection{Angular and distance distributions in 2D networks}

\begin{figure}[t]
    \centering
    \includegraphics[width=0.65\columnwidth]{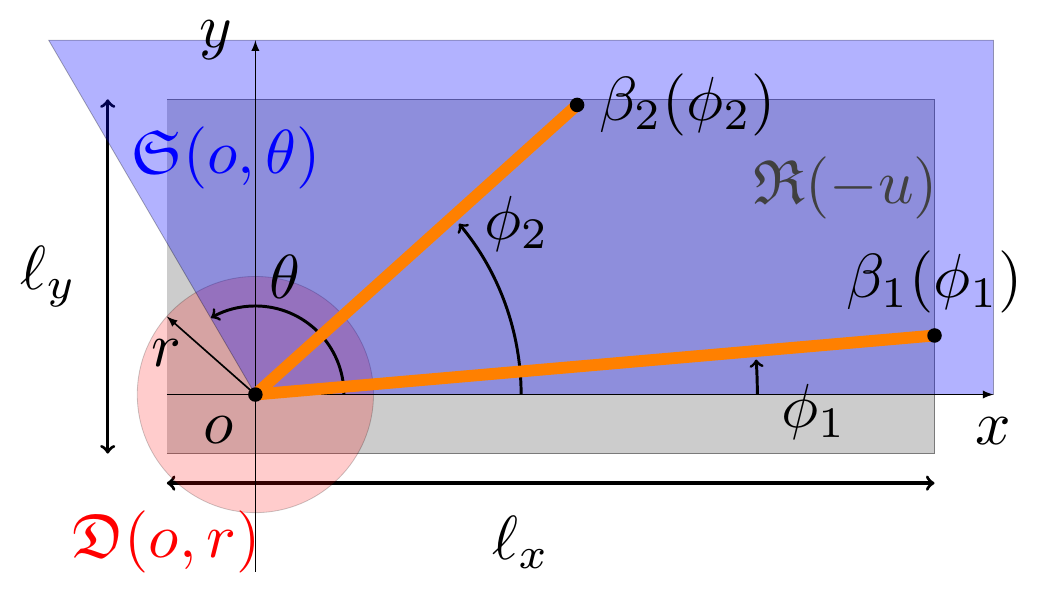}
    \caption{{Sketch of the relevant sets involved in the derivation of the joint distribution of distance and angle within a rectangular region.}}
    \label{fig:rectangularRegion}
\end{figure}

Let us assume a rectangular region, $\mathfrak{R}(o)$, centered at the origin $o$, whose side lengths are $\ell_x$ and $\ell_y$ on the $x$ and $y$ axis, respectively {(see Fig.} \ref{fig:rectangularRegion}{)}. We aim at deriving the joint distribution of the distance and azimuth angle for the link between a random point 
and a reference point, $u=(u_x,u_y)$, with $\lvert u_x \rvert < \frac{\ell_x}{2} \wedge \lvert u_y \rvert < \frac{\ell_y}{2}$. As pointed out in Section \ref{sec:ArbitraryShaped}, this is equivalent to computing the overlap area of a disk and a sector after translating the regions by $-u$. To that end, we follow the 3 steps provided in Section \ref{sec:ArbitraryShaped}, starting with the derivation of $\beta (\phi)$, given in the following lemma.

\begin{lemma}[\textbf{Step 1}]
    \label{lem:Rectangle_in_polar}
    A rectangular region of sides $\ell_x$ and $\ell_y$ centered at $-u$, with $\lvert u_x \rvert < \frac{\ell_x}{2} \wedge \lvert u_y \rvert < \frac{\ell_y}{2}$,  can be expressed in polar coordinates as 
    \begin{equation}
        \label{eq:Convex_region_in_polar}
        \mathfrak{R}(-u) = \{ \rho > 0, \phi \in [0, 2\pi] \lvert \mathcal{B}(\rho,\phi) \},
    \end{equation}
    \noindent where $\mathcal{B}(\rho,\phi) = \{\rho \leq \beta(\phi)\}$ and 
    \begin{equation}
        \beta(\phi) = \min \left( \frac{h_x(\phi)}{\cos(\phi)}, \frac{h_y(\phi)}{\sin(\phi)} \right),
    \end{equation}
    \noindent where
    \begin{align}
        h_x(\phi) &= h_{x^+} \mathbbm{1} \left( \mathcal{A}_c (\phi) \right) + h_{x^-} \mathbbm{1} \left( \bar{\mathcal{A}_c} (\phi) \right), \nonumber \\
        h_y(\phi) &= h_{y^+} \mathbbm{1} \left( \mathcal{A}_s (\phi) \right) + h_{y^-} \mathbbm{1} \left( \bar{\mathcal{A}_s} (\phi) \right), 
    \end{align} 
    \noindent with $h_{x^-} = -\frac{\ell_x}{2} - u_x$, $h_{x^+} = \frac{\ell_x}{2} - u_x$, 
    $h_{y^-} = -\frac{\ell_y}{2} - u_y$, $h_{y^+} = \frac{\ell_y}{2} - u_y$, and
    \begin{align}
        \mathcal{A}_c (\phi) = \left( \phi \in \mathfrak{Q}_{1} \right) \vee \left(\phi \in \mathfrak{Q}_{4} \right); \, 
        \bar{\mathcal{A}_c} (\phi) = \left( \phi \in \mathfrak{Q}_{2} \right) \vee \left(\phi \in \mathfrak{Q}_{3} \right), \nonumber \\
        \mathcal{A}_s (\phi) = \left( \phi \in \mathfrak{Q}_{1} \right) \vee \left(\phi \in \mathfrak{Q}_{2} \right); \,
        \bar{\mathcal{A}_s} (\phi) = \left( \phi \in \mathfrak{Q}_{3} \right) \vee \left(\phi \in \mathfrak{Q}_{4} \right),
    \end{align}
    \noindent where $\mathfrak{Q}_{l}$, $l = \{1, 2, 3, 4\}$, represent the $4$ angular quadrants as $\mathfrak{Q}_{l} = \left[\frac{ (l-1)\pi}{2},  \frac{l \pi}{2} \right)$.
    \begin{proof}
        See Appendix \ref{App:Proof of lem:Rectangle_in_polar}.
    \end{proof}
\end{lemma}

The next lemma gives the sets $\mathfrak{T_{\mathfrak{B}}}(r)$ and $\mathfrak{T_{\bar{\mathfrak{B}}}}(r)$, following \textbf{Step 2} of the proposed procedure. 

\begin{lemma}[\textbf{Step 2}]
    \label{lem:Step2}
    The sets $\mathfrak{T_{\mathfrak{B}}}(r)$ and $\mathfrak{T_{\bar{\mathfrak{B}}}}(r)$ are expressed as the union of $8$ disjoint sets as
    \begin{align}
        \mathfrak{T_{\mathfrak{B}}}(r) = \bigcup_{i=1}^8 \mathfrak{X}_i(r); \;  \mathfrak{T_{\bar{\mathfrak{B}}}}(r) = \bigcup_{i=1}^8 \mathfrak{M}_i(r).
    \end{align}
    \noindent with 
    \begin{equation}
        \label{eq:X_i}
        \mathfrak{X}_i(r) = 
            \begin{cases}
                \left[\chi^{(\geq)}_{i,1}(r), \chi^{(\geq)}_{i,2}(r) \right), & \mathrm{if} \; r \geq h_i, \\            
                \left[\chi^{(<)}_{i,1}(r), \chi^{(<)}_{i,2}(r) \right), & \mathrm{if} \; r < h_i,
            \end{cases}
    \end{equation}
    \noindent and
    \begin{equation}
        \label{eq:M_i}
    \mathfrak{M}_i(r) = 
        \begin{cases}
            \left[\mu_{i,1}(r), \mu_{i,2}(r) \right), & \mathrm{if} \; r \geq h_i, \\          
            \emptyset, & \mathrm{if} \; r < h_i,
        \end{cases}
    \end{equation}
    \noindent where $\chi^{(\geq)}_{i,j}(r), \chi^{(<)}_{i,j}(r)$, and $\mu_{i,j}(r)$, $i=\{1,\ldots,8\}$, $j=\{1,2\}$, are given in \eqref{eq:chi_i(r)} and \eqref{eq:mu_i(r)}, and $[h_1, h_2, h_3, h_4, h_5, h_6, h_8] = [h_{x^+}, h_{x^+}, -h_{x^-}, -h_{x^-}, h_{y^+}, h_{y^+}, -h_{y^-}, -h_{y^-}]$ .
    \begin{figure*}[t]
        \begin{align}
            \label{eq:chi_i(r)}
            \chi^{(<)}_{1,1} &= 0,  &
            \chi^{(<)}_{1,2} &= \atan(\frac{h_{y^+}}{h_{x^+}}),  &
            \chi^{(\geq)}_{1,1}(r) &= \acos\left(\frac{h_{x^+}}{r} \right),  &
            \chi^{(\geq)}_{1,2}(r) &= \atan\left(\frac{h_{y^+}}{h_{x^-}} \right), \nonumber \\
            \chi^{(<)}_{2,1} &= \atan(\frac{h_{y^-}}{h_{x^+}}) + 2\pi,  &
            \chi^{(<)}_{2,2} &= 2\pi, &
            \chi^{(\geq)}_{2,1}(r) &= \atan\left(\frac{h_{y^-}}{h_{x^+}} \right) + 2\pi, &
            \chi^{(\geq)}_{2,2}(r) &=  2\pi - \acos\left(\frac{h_{x^+}}{r} \right), \nonumber \\
            \chi^{(<)}_{3,1} &= \pi, &
            \chi^{(<)}_{3,2} &= \atan(\frac{h_{y^-}}{h_{x^-}}) + \pi, &
            \chi^{(\geq)}_{3,1}(r) &= 2\pi - \acos\left(\frac{h_{x^-}}{r} \right), &
            \chi^{(\geq)}_{3,2}(r) &= \atan\left(\frac{h_{y^-}}{h_{x^-}} \right) + \pi, \nonumber \\
            \chi^{(<)}_{4,1} &= \atan(\frac{h_{y^+}}{h_{x^-}}) + \pi, &
            \chi^{(<)}_{4,2} &= \pi, &
            \chi^{(\geq)}_{4,1}(r) &= \atan\left(\frac{h_{y^+}}{h_{x^-}} \right) + \pi, &
            \chi^{(\geq)}_{4,2}(r) &=  \acos\left(\frac{h_{x^-}}{r} \right),\nonumber \\
            \chi^{(<)}_{5,1} &= \atan(\frac{h_{y^+}}{h_{x^+}}), &
            \chi^{(<)}_{5,2} &= \frac{\pi}{2}, &
            \chi^{(\geq)}_{5,1}(r) &= \atan\left(\frac{h_{y^+}}{h_{x^+}} \right), &
            \chi^{(\geq)}_{5,2}(r) &= \asin\left(\frac{h_{y^+}}{r} \right), \nonumber \\
            \chi^{(<)}_{6,1} &= \frac{\pi}{2}, &
            \chi^{(<)}_{6,2} &= \atan(\frac{h_{y^+}}{h_{x^-}}) + \pi, &
            \chi^{(\geq)}_{6,1}(r) &= \pi - \asin\left(\frac{h_{y^+}}{r} \right), &
            \chi^{(\geq)}_{6,2}(r) &= \atan\left(\frac{h_{y^+}}{h_{x^-}} \right) +\pi, \nonumber \\
            \chi^{(<)}_{7,1} &= \atan(\frac{h_{y^-}}{h_{x^-}}) +\pi, &
            \chi^{(<)}_{7,2} &= \frac{3\pi}{2}, &
            \chi^{(\geq)}_{7,1}(r) &= \atan\left(\frac{h_{y^-}}{h_{x^-}} \right) + \pi, &
            \chi^{(\geq)}_{7,2}(r) &=  \pi - \asin\left(\frac{h_{y^-}}{r} \right), \nonumber \\
            \chi^{(<)}_{8,1} &= \frac{3\pi}{2}, &
            \chi^{(<)}_{8,2} &= \atan(\frac{h_{y^-}}{h_{x^+}}) + 2\pi, &
            \chi^{(\geq)}_{8,1}(r) &= 2\pi + \asin\left(\frac{h_{y^-}}{r} \right), &
            \chi^{(\geq)}_{8,2}(r) &= \atan(\frac{h_{y^-}}{h_{x^+}}) + 2\pi. \nonumber \\
        \end{align}
        \hrulefill
%
        \begin{align}
            \label{eq:mu_i(r)}
            \mu_{1,1}(r) &=  0,&
            \mu_{1,2}(r) &=  \min\left(\acos(\frac{h_{x^+}}{r}), \atan(\frac{h_{y^+}}{h_{x^+}}) \right), \nonumber \\
            \mu_{2,1}(r) &=  \max\left(2\pi - \acos\left(\frac{h_{x^+}}{r} \right), \atan\left(\frac{h_{y^-}}{h_{x^+}} \right) + 2\pi \right),&
            \mu_{2,2}(r) &=  2\pi, \nonumber \\
            \mu_{3,1}(r) &=  \pi,&
            \mu_{3,2}(r) &=  \min\left(2\pi - \acos\left(\frac{h_{x^-}}{r} \right), \atan\left(\frac{h_{y^-}}{h_{x^-}} \right) + \pi \right),\nonumber \\
            \mu_{4,1}(r) &=  \max\left( \atan\left(\frac{h_{y^+}}{h_{x^-}} \right) + \pi, \acos\left(\frac{h_{x^-}}{r} \right) \right),&
            \mu_{4,2}(r) &=   \pi, \nonumber \\
            \mu_{5,1}(r) &=  \max\left(  \atan\left(\frac{h_{y^+}}{h_{x^+}} \right), \asin\left(\frac{h_{y^+}}{r} \right) \right),&
            \mu_{5,2}(r) &=  \frac{\pi}{2}, \nonumber \\
            \mu_{6,1}(r) &=  \frac{\pi}{2},&
            \mu_{6,2}(r) &=  \min\left(\pi - \asin\left(\frac{h_{y^+}}{r} \right), \atan\left(\frac{h_{y^+}}{h_{x^-}} \right) +\pi \right), \nonumber \\
            \mu_{7,1}(r) &=  \max\left(\atan\left(\frac{h_{y^-}}{h_{x^-}} \right) + \pi, \pi - \asin\left(\frac{h_{y^-}}{r} \right) \right),&
            \mu_{7,2}(r) &=  \frac{3\pi}{2},\nonumber \\
            \mu_{8,1}(r) &=  \frac{3\pi}{2},&
            \mu_{8,2}(r) &=   \min\left(2\pi + \asin\left(\frac{h_{y^-}}{r} \right), \atan(\frac{h_{y^-}}{h_{x^+}}) + 2\pi \right) .
        \end{align}
        \hrulefill
    \end{figure*}
    \begin{proof}
        See Appendix \ref{App:Proof of lem:Step2}.
    \end{proof}
\end{lemma}


\begin{corollary}
    \label{cor:Disjoint_set}
    The subsets $\mathfrak{X}_i(r)$ and $\mathfrak{M}_i(r)$ in \eqref{eq:X_i} and \eqref{eq:M_i} are mutually disjoint, i.e.,
    \begin{align}
        \bigcap_{i=1}^8 \mathfrak{X}_i(r) = \bigcap_{i=1}^8 \mathfrak{M}_i(r) = \emptyset.
    \end{align} 
    Each subset
    is restricted to a given quadrant as follows:
    \begin{align}
        \mathfrak{X}_1(r) &\subset \mathfrak{Q}_1, & \mathfrak{X}_2(r) &\subset \mathfrak{Q}_4, & \mathfrak{X}_3(r) &\subset \mathfrak{Q}_3, & \mathfrak{X}_4(r) &\subset \mathfrak{Q}_2, \nonumber \\
        \mathfrak{X}_5(r) &\subset \mathfrak{Q}_1, & \mathfrak{X}_6(r) &\subset \mathfrak{Q}_2, & \mathfrak{X}_7(r) &\subset \mathfrak{Q}_3, & \mathfrak{X}_8(r) &\subset \mathfrak{Q}_4, \nonumber \\
        \mathfrak{M}_1(r) &\subset \mathfrak{Q}_1, & \mathfrak{M}_2(r) &\subset \mathfrak{Q}_4, & \mathfrak{M}_3(r) &\subset \mathfrak{Q}_3, & \mathfrak{M}_4(r) &\subset \mathfrak{Q}_2, \nonumber \\
        \mathfrak{M}_5(r) &\subset \mathfrak{Q}_1, & \mathfrak{M}_6(r) &\subset \mathfrak{Q}_2, & \mathfrak{M}_7(r) &\subset \mathfrak{Q}_3, & \mathfrak{M}_8(r)   &\subset \mathfrak{Q}_4.
    \end{align}     
    \begin{proof}
        See Appendix \ref{App:Proof of cor:Disjoint_set}.
    \end{proof}
\end{corollary}

It only remains to complete \textbf{step 3} of the framework to derive the joint CDF of the distance and angle, given next. 

\begin{theorem}
\label{theor:Joint CDF}
The joint CDF of the distance and angle of random points, uniformly distributed in a rectangle $\mathfrak{R}(o)$, towards a reference point $u=(u_x, u_y)$, can be written as


\begin{align}
    \label{eq:Joint_CDF}
F_{R,\Theta} &(r, \theta) =  \nonumber \\
&\frac{r^2}{2 \ell_x \ell_y}\Bigg[  \sum_{i=1}^{8} \mathbbm{1} (r < h_i) \Bigg(\min \Big(\theta, \chi^{(<)}_{i,2}(r)\Big) - \chi^{(<)}_{i,1}(r) \Bigg)^+ \nonumber \\
& + \mathbbm{1} (r \geq h_i) \Bigg(\min \Big(\theta, \chi^{(\geq)}_{i,2}(r)\Big) - \chi^{(\geq)}_{i,1}(r) \Bigg)^+ \Bigg] \nonumber \\
& +  \sum_{i=1}^{4}\frac{h_i^2}{2 \ell_x \ell_y} \mathbbm{1} (r \geq h_i) \mathbbm{F} \Big(\tan; \mu_{i,1}(r); \min(\theta, \mu_{i,2}(r)) \Big) \nonumber \\
&-  \sum_{i=5}^{8}\frac{h_i^2}{2 \ell_x \ell_y} \mathbbm{1} (r \geq h_i) \mathbbm{F} \Big(\tan^{-1}; \mu_{i,1}(r); \min(\theta, \mu_{i,2}(r))\Big),
\end{align}
\noindent with $h_i$, $\chi^{(<)}_{i,j}(r)$, $\chi^{(\geq)}_{i,j}(r)$ and $\mu_{i,j}(r)$ given in \textbf{Lemma \ref{lem:Step2}} with \eqref{eq:chi_i(r)} and \eqref{eq:mu_i(r)} for $i=\{1,\ldots,8\}$, $j=\{1,2\}$. 
\begin{proof}
    See Appendix \ref{App:Proof of theor:Joint CDF}.
\end{proof}
\end{theorem} 

The joint CDF of \textbf{Theorem \ref{theor:Joint CDF}} is given in closed-form as the sum of $16$ simple terms involving compositions of trigonometric and $\min(x,y)$ functions. We now derive the joint PDF and the marginals for the azimuth angle.  

\begin{corollary} In the settings of Theorem \ref{theor:Joint CDF}, the joint PDF of the distance and angle is given by
    \label{cor:joint_pdf_R_Theta}
    \begin{align}
        f_{R,\Theta} (r, \theta) = \frac{r}{\ell_x \ell_y}  \sum_{i=1}^{8} &\mathbbm{1} (r < h_i) 
            \mathbbm{1}_{\Big[\chi^{(<)}_{i,1}, \chi^{(<)}_{i,2}\Big)}(\theta) \nonumber \\
        + &\mathbbm{1} (r \geq h_i) \mathbbm{1}_{\Big[\chi^{(\geq)}_{i,1}(r), \chi^{(\geq)}_{i,2}(r)\Big)}(\theta).
    \end{align}
    %
    %
    \begin{proof}
See Appendix \ref{App:Proof_corollaries}.      
    \end{proof}
\end{corollary}


\begin{corollary}
    \label{cor:marginal_azimuth_CDF}
    The marginal CDF of the angle between a reference point at $u=(u_x, u_y)$ and uniformly distributed random points placed within the rectangle $\mathfrak{R}(o)$ is given by
    %
    \begin{align}
        F_\Theta (\theta) = \frac{1}{2\ell_x \ell_y} \Bigg( &\sum_{i=1}^{4} h_i^2 \mathbbm{F} (\tan; \epsilon_{i,1}(\theta); \epsilon_{i,2}(\theta)) \nonumber \\
        - &\sum_{i=5}^{8} h_i^2 \mathbbm{F} (\tan^{-1};  \epsilon_{i,1}(\theta);  \epsilon_{i,2}(\theta)) \Bigg),
    \end{align}
    \noindent where $h_i$ for $i=\{1,\ldots,8\}$ is given in \textbf{Lemma \ref{lem:Step2}} and 
    \begin{align}
        \epsilon_{1,1}(\theta), &= 0 & \epsilon_{1,2}(\theta) &= \min \left(\theta, \atan(\frac{h_{y^+}}{h_{x^+}})\right), & \nonumber \\
        \epsilon_{2,1}(\theta) &= \atan(\frac{h_{y^-}}{h_{x^+}}) + 2\pi, & \epsilon_{2,2}(\theta) &= \theta, & \nonumber 
    \end{align}
    %

    %
    \begin{align}
        \epsilon_{3,1}(\theta) &= \pi, & \epsilon_{3,2}(\theta) &= \min\left( \theta, \atan(\frac{h_{y^-}}{h_{x^-}}) + \pi \right), & \nonumber \\
        \epsilon_{4,1}(\theta) &= \atan(\frac{h_{y^+}}{h_{x^-}}) + \pi, & \epsilon_{4,2}(\theta) &= \min(\theta, \pi), & \nonumber \\
        \epsilon_{5,1}(\theta) &= \atan(\frac{h_{y^+}}{h_{x^+}}), & \epsilon_{5,2}(\theta) &= \min \left(\theta, \frac{\pi}{2}\right), & \nonumber \\
        \epsilon_{6,1}(\theta) &= \frac{\pi}{2}, & \epsilon_{6,2}(\theta) &= \atan(\frac{h_{y^+}}{h_{x^-}}) + \pi, & \nonumber \\
        \epsilon_{7,1}(\theta) &= \atan(\frac{h_{y^-}}{h_{x^-}}) + \pi, & \epsilon_{7,2}(\theta) &= \min \left(\theta, \frac{3\pi}{2}\right), & \nonumber \\
        \epsilon_{8,1}(\theta) &= \frac{3\pi}{2} & \epsilon_{8,2}(\theta) &= \atan(\frac{h_{y^-}}{h_{x^+}}) + 2\pi. & \nonumber 
    \end{align}
    \begin{proof}
         See Appendix \ref{App:Proof_corollaries}. 
    \end{proof}
\end{corollary}

\begin{corollary}
    \label{cor:marginal_azimuth_PDF}
    The marginal PDF of the azimuth angle, $\Theta$, can be expressed as
    \begin{align}
        f_\Theta (\theta) = \frac{1}{2\ell_x \ell_y} \Bigg( &\sum_{i=1}^{4} \frac{h_i^2}{\cos^2(\theta)} \mathbbm{1}_{\big[\chi^{(<)}_{i,1},\chi^{(<)}_{i,2}\big)}(\theta) \nonumber \\
        + &\sum_{i=5}^{8} \frac{h_i^2}{\sin^2(\theta)} \mathbbm{1}_{\big[\chi^{(<)}_{i,1},\chi^{(<)}_{i,1}\big)}(\theta) \Bigg),
    \end{align}
    \noindent where $h_i$ for $i=\{1,\ldots,8\}$ are given in \textbf{Lemma \ref{lem:Step2}} and the terms $\chi^{(<)}_{i,j}$ with $j\in\{1,2\}$ are given in \label{eq:chi_i(r)}. 
    {It is important to remark that the intervals $\big[\chi^{(<)}_{i,1},\chi^{(<)}_{i,2}\big)$ are disjoint sets for $i\in \{1,..,8\}$.}
    \begin{proof}
        The result is readily obtained from the derivative of $F_\theta(\theta)$, tanking into account \textbf{Definition \ref{def:F operator}}. 
    \end{proof}
\end{corollary}

\subsection{Asymptotic case: $\ell_y \gg \ell_x$}
{In this section we investigate the case when one of the sides of a rectangle is much greater than the other. Without loss of generality, we consider that $\ell_y \gg \ell_x$ and the area of the rectangle is $1$. We formulate mathematically this case as $\ell_y \to \infty$, with $\ell_y \ell_x = 1$, i.e., $\ell_x \to 0^+$, and $\lvert u \rvert < \infty$. The following corollary gives the marginal distribution of the azimuth angle in such case.} 
\begin{corollary}
    {The marginal distribution of the azimuth angle, $\theta$, when $\ell_y \to \infty$, with $\ell_y \ell_x = 1$ can be expressed as follows}:
    \textcolor{black}{\begin{equation}
        \lim_{\ell_y \to \infty} f_\Theta(\theta) = \frac{1}{2} \left(\delta\left(\theta 
            - \frac{\pi}{2}\right) + \delta\left(\theta - \frac{3\pi}{2}\right) \right),
    \end{equation}}
    \noindent {where $\delta(x)$ is the Dirac Delta function}. 
    \begin{proof}
        {The proof comes after realizing that the first $4$ terms of the PDF given with \mbox{\textbf{Corollary \ref{cor:marginal_azimuth_PDF}}} are $0$ since $h_{x^+}$ and $h_{x^-}$ are $0$ when $\ell_y \to \infty$, i.e., $\ell_x \to 0^+$. Manipulating the remaining terms for $i=\{5,..,8\}$, and grouping them leads to:}
        \textcolor{black}{\begin{align}
            \label{eq:asympotic_1}
            \lim_{\ell_y \to \infty} &f_\Theta(\theta) = \frac{1}{2} \lim_{\ell_y \to \infty} \overbrace{\frac{h^2_{y^+}}{\sin(\theta)^2} \mathbbm{1}_{\left[\mathrm{atan}\left(\frac{h_{y^+}}{h_{x^+}}\right), \mathrm{atan}\left(\frac{h_{y^+}}{h_{x^-}}\right) + \pi \right)} (\theta)}^{g(\theta - \pi/2, \ell_y)} \nonumber \\
            &+ \frac{1}{2} \lim_{\ell_y \to \infty} \underbrace{\frac{h^2_{y^-}}{\sin(\theta)^2} \mathbbm{1}_{\left[\mathrm{atan}\left(\frac{h_{y^-}}{h_{x^-}}\right) + \pi, \mathrm{atan}\left(\frac{h_{y^-}}{h_{x^+}}\right) + 2\pi \right)} (\theta)}_{g(\theta - 3\pi/2, \ell_y)}.
        \end{align}}
        {The next step is to prove that both terms fulfills the two conditions imposed over the Dirac Delta function as per \textbf{Definition \mbox{\ref{def:DiracDelta}}}. 
        As $\ell_y$ increases the indicator function, which can be viewed as a pulse on $\theta$, gets narrower and centered around $[\pi^-/2, \pi^+/2)$ for the first term, and $[3\pi^-/2, 3\pi^+/2)$, for the second term, when $\ell_y\to \infty$. On the other, the amplitude of the terms ${h^2_{y^+}}/{\sin(\theta)^2}$ and ${h^2_{y^+}}/{\sin(\theta)^2}$ tend to infinity in the limit case. Lastly, identifying the functions $g(\theta - \pi/2, \ell_y)$ and $g(\theta - 3\pi/2, \ell_y)$ in \mbox{\eqref{eq:asympotic_1}}, and checking that they have unit integral over its domain and tend to infinity on $\pi/2$ and $3\pi/2$ while being $0$ outside completes the proof.}
    \end{proof}
\end{corollary}
{The above corollary is a formal proof that the angles of a rectangle with one of the sides much greater than the other tend to be concentrated on two possible values, pointing to the directions of maximal length, i.e., $\pi/2$ and $3\pi/2$ when $\ell_y \to \infty$. 
To preserve a finite area, increasing one of the sides involves a reduction of the other, and thus, in the limit case with $\ell_y \to \infty, \ell_x \to 0$, the nodes are randomly distributed within a 1D line. This motivates the following approximation to the joint PDF of distances and angles.}  
\begin{approximation}
\label{appr:joint PDF}
    {The joint PDF of distances and angles when $\ell_y \gg \ell_x$ and the reference node, $u$, is placed at the center of the rectangle can be approximated as:}
    \textcolor{black}{\begin{equation}
        f_{R, \Theta} (r, \theta) \approx f_R(r) \left( \frac{1}{2} \delta\left(\theta - \frac{\pi}{2} \right) + \frac{1}{2} \delta\left(\theta - \frac{3\pi}{2} \right) \right),
    \end{equation}}
    \noindent { where $f_R(r)$ stands for the distribution of distances in a 1D line of length $\ell_y$, which is written as $f_R(r) = 2/\ell_y, r\leq\ell_y$, and $0$ otherwise.}
\end{approximation}

\subsection{Extension to 3D Networks: impact of height}

For the 3D case, we consider an arbitrary reference point placed at $u=(u_x,u_y,u_z) \in \mathbb{R}^3$ and random points $V=(V_x, V_y, v_z)$ with fixed (deterministic) height $v_z$. When projected on the $xy$ plane, the reference point $u$ falls within a 2D convex region $\mathfrak{R}(o)$, with {the} center of mass located at the origin. Similarly, the projection (on the $xy$ plane) of random points $V$ lies in $\mathfrak{R}(o)$, i.e., $(V_x, V_y) \in \mathfrak{R}(o)$. 
%
%
This is a realistic scenario in many applications. 
{For instance, in terrestrial wireless networks, while users are typically modeled with random locations, their antennas are assumed to have a deterministic (fixed) height--since such height is very similar for all users; at the other end, the height of access points (AP) or base station (BS) antennas is an important design parameter, which is relevant in the design of directional beamforming (e.g., design of analog codebooks) through the joint azimuth and zenith angle dependence. In aerial networks such as UAV-based networks, the height of the UVAs and the BSs--which provide the backhaul links--are considered deterministic} \cite{Banagar21}{, and this is a parameter which needs to be optimized in the network design.}



We are interested {in} the joint distribution of the distance, azimuth and zenith angles of the link between $u$ and $V$, given in the next theorem.



\begin{theorem}
    \label{theo:Joint_CDF_3D}
    The joint distribution of distance ($D$), azimuth ($\Theta$) and zenith ($\Psi$) angles for the link between a reference point at $u=(u_x,u_y,u_z) \in \mathbb{R}^3$ and random points $V=(V_x, V_y, v_z)$, with $(V_x, V_y)$ uniformly distributed in a convex region $\mathfrak{R}(o) \subset \mathbb{R}^2$, and deterministic $v_z$, is expressed as {given with \mbox{\eqref{eq:3D_CDF}}}
    
    
    \begin{figure*}[t]
        \begin{align}
        \label{eq:3D_CDF}
        & F_{D, \Theta, \Psi} (\mathscr{d}, \theta, \psi) = \mathbbm{1}(\mathscr{d} > \lvert u_z-v_z \lvert \wedge u_z \geq v_z) 
            \Bigg[\mathbbm{1}_{\mathfrak{Q}_{2}}(\psi) 
            \Bigg(F_{R, \Theta}\left(\sqrt{d^2 - (u_z-v_z)^2}, \theta\right) - F_{R, \Theta}\Big((u_z-v_z) \tan(\pi-\psi), \theta\Big)\Bigg)^+ \\
        & \; + \mathbbm{1}_{\mathfrak{Q}_{1}}(\psi) 
            F_{R, \Theta}\left(\sqrt{d^2 - (u_z-v_z)^2}, \theta\right)\Bigg]
        + \mathbbm{1}(\mathscr{d} > \lvert u_z-v_z \lvert \wedge u_z < v_z) \mathbbm{1}_{\mathfrak{Q}_{1}}(\psi) 
        F_{R, \Theta}\left(\min\left(\sqrt{d^2 - (u_z-v_z)^2}, (v_z-u_z) \tan(\psi)\right), \theta\right) \nonumber
        \end{align}
        \hrulefill
    \end{figure*}
    
    \noindent where $F_{R, \Theta}(r,\theta)$ is the joint distribution of distance and azimuth angle in the $xy$ plane, given in \textbf{Theorem
    \ref{theor:Joint CDF}}.
    \begin{proof}
    See Appendix \ref{App:Proof of theo:Joint_CDF_3D}.
    \end{proof}
\end{theorem}

In \textbf{Theorem \ref{theo:Joint_CDF_3D}}, as expected, the CDF is $0$ when $\mathscr{d} < \lvert u_z-v_z \lvert$, because the distance between the random nodes and the reference node should be greater than the difference of their antenna heights. 
This theorem enables the analysis of performance metrics of interest, {e.g.,} the coverage probability, in finite networks where the height of the antennas is relevant. 
Other joint statistics relating to the angular domain are also instrumental. In particular, the joint distribution of azimuth and zenith angles can be exploited for the design of analog codebooks. We characterize this distribution in the following two corollaries, providing the joint CDF and PDF. 

\begin{corollary}
    \label{cor:3D_joint_angular_CDF}
    The joint CDF of azimuth ($\Theta$) and zenith ($\Psi$) angles is given by
    \begin{align}
        & F_{\Theta, \Psi} (\theta, \psi) = \mathbbm{1}(u_z\geq v_z) \Bigg[ \mathbbm{1}\left(\psi \in \left[\frac{\pi}{2}, \pi\right] \right) 
        \Big(F_\Theta(\theta) 
        \\ 
        & \quad - F_{R, \Theta} ((u_z-v_z) \tan(\pi - \psi), \theta) \Big)^+ + \mathbbm{1}\left(\psi \in \left[0,\frac{\pi}{2}\right) \right) F_\Theta(\theta) \Bigg]
        \nonumber \\
        & \quad + \mathbbm{1}(u_z < v_z)  \mathbbm{1}\left(\psi \in \left[0,\frac{\pi}{2}\right) \right) F_{R, \Theta} ((v_z-u_z) \tan(\psi), \theta). \nonumber
    \end{align}
    \begin{proof}
        We arrive at the result after computing the limit $F_{\Theta, \Psi}(\theta, \psi) = \lim \limits_{\mathscr{d}\to\infty} F_{D, \Theta, \Psi}(\mathscr{d}, \theta, \psi)$, with $F_{D, \Theta, \Psi}(\mathscr{d}, \theta, \psi)$ given in \textbf{Theorem \ref{theo:Joint_CDF_3D}}. 
    \end{proof}
\end{corollary}

\begin{corollary}
    \label{cor:3D_joint_angular_PDF}
    The joint PDF of azimuth ($\Theta$) and zenith ($\Psi$) angles is given by
    \begin{align}
        & f_{\Theta, \Psi}(\theta, \psi) = \mathbbm{1}(u_z < v_z) \mathbbm{1}\left(\psi \in \left[0, \frac{\pi}{2}\right) \right) 
        \frac{v_z-u_z}{\cos^2(\psi)} \nonumber \\ 
        & \quad f_{R, \Theta}((v_z-u_z) \tan(\psi), \theta) - \mathbbm{1}(u_z \geq v_z) \mathbbm{1}\left(\psi \in \left[\frac{\pi}{2}, \pi\right] \right) 
        \nonumber \\
        & \quad f_{R, \Theta}((u_z-v_z) \tan(\pi-\psi), \theta) \frac{v_z-u_z}{\cos^2(\psi)}. 
    \end{align}
    \begin{proof}
        The result is obtained from the partial derivatives of the joint CDF (given in \textbf{Corollary \ref{cor:3D_joint_angular_CDF}}) with respect to the azimuth and zenith angles, i.e., $f_{\Theta, \Psi}(\theta, \psi) = \pdv{F_{\Theta, \Psi}(\theta, \psi)}{\theta}{\psi}$.
    \end{proof}
\end{corollary}

\section{Applications}
\label{sec:applications}
\begin{figure}[t]
    \centering
    \includegraphics[width=\columnwidth]{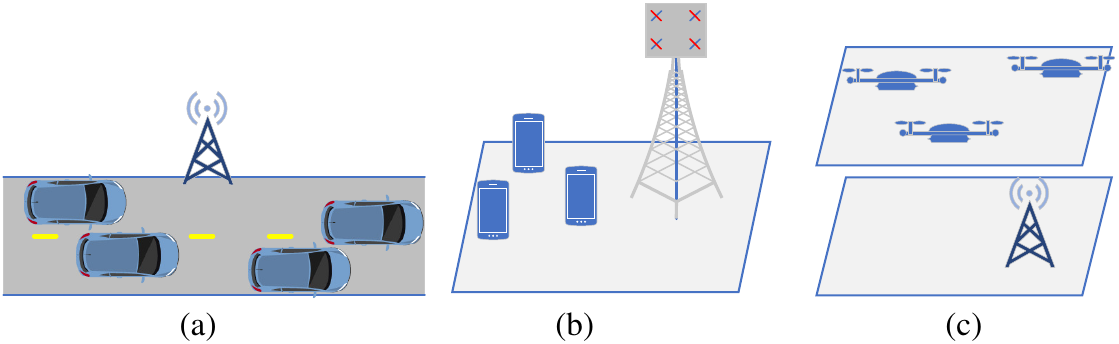}
    \caption{Sketch that illustrates examples of different scenarios that can be modeled with the proposed framework: (a) vehicular communications; (b) indoor/outdoor wireless systems; (c) terrestrial-to-aerial networks. }
    \label{fig:scenarios}
\end{figure}

The proposed mathematical framework allows for modeling {many} scenarios of practical relevance in wireless communications. As illustrated in Fig. \ref{fig:scenarios}, a rectangle with $\ell_x \gg \ell_y$ can model the road to consider the case of vehicular communications. In this case, a mounted AP can be considered the reference node (placed at $u$) that communicates with vehicles randomly located along the road. The cases of indoor WiFi-based communications or outdoor cellular communications can also be also considered{; e.g.,} for terrestrial communications, a BS with {a} given antenna height at a reference location {and} a hot-spot of users with smaller antenna height; in indoor scenarios, the reference node could be an AP mounted on the ceiling. Aerial-to-terrestrial communications {(}e.g., based on UAVs{)} can also be considered; for instance, a BS located at a reference position $u$, {providing} backhaul access to a network of UAVs {that fly} at a given altitude, much higher than the BS. 

A rectangular region is an appropriate modeling choice for these aforementioned scenarios, where our results can be applied.
More specifically, the joint distance and angle distribution given in \textbf{Theorem \ref{theor:Joint CDF}} and \textbf{\ref{theo:Joint_CDF_3D}} are needed to compute the distribution of the SNR when users are placed within a finite area{,} and directional radiation patterns are used. In the general 3D case, the SNR for the link between a transmit node placed at $u$ and a randomly located node can be expressed as 
\begin{equation}
    \label{eq:SNR}
    \mathrm{SNR} = \frac{g_{\rm t}\left(\Theta_t, \Psi_t\right) g_{\rm r}\left(\Theta_r, \Psi_r\right) (\tau D)^{-\alpha} \lvert \beta \lvert^2 \rho_t}{N_0},
\end{equation}
\noindent where $\Theta_t, \Psi_t$ and $\Theta_r, \Psi_r$ are the transmit and receive azimuth and zenith angles, respectively; 
$g_{\rm t}(\bullet)$ and $g_{\rm r}(\bullet)$ represent the transmit and receive antenna gains (radiation patterns); $\lvert \beta \lvert$ is the fast-fading amplitude; $\tau,\alpha$ are the path loss slope and exponent; and $\rho_t, N_0$ are the transmit power spectral density (PSD) and noise PSD, respectively. If we consider the downlink, with a BS placed at $u$ and randomly positioned nodes, the joint distribution of \textbf{Theorem \ref{theo:Joint_CDF_3D}} would model the transmit angles $\Theta_t, \Psi_t$ and the distance $D$, whereas the receive angles, $\Theta_r, \Psi_r$, would be obtained from the transmit angles after simple trigonometric transformations. Importantly, our results do not make any assumption on the radiation patterns, as opposed to previous works restricted to a sector model, e.g., \cite{Azimi-Abarghouyi19}, or assuming perfect beam alignment, e.g., \cite{DaSilva20,Kokkoniemi20}. Our results hold for any radiation patterns $g_{\rm t}(\bullet)$ and $g_{\rm r}(\bullet)$, which can be related either to single-element antennas, e.g., horn antennas \cite{Balanis12}, or antenna arrays like uniform planar arrays (UPAs). For the latter, to compute the SNR in \eqref{eq:SNR}, the product of gain patterns, $g_{\rm t}(\bullet) g_{\rm r}(\bullet)$, should be replaced by the product of {transmit and received beamforming matrices (i.e., $\mathbf{w}_t$ and $\mathbf{w}_r$ respectively) by their array response vectors, $\lvert \mathbf{a_t}\left(\Theta_t, \Psi_t\right)^H \mathbf{w_t} \lvert^2 \lvert \mathbf{a_r}\left(\Theta_r, \Psi_r\right)^H \mathbf{w_r} \lvert^2$}, \cite{Bjornson17}.
%
Therefore, our results can be applied to the analysis of emerging techniques related to directional beamforming, such as beam management procedures in 5G and beyond, and the analysis of RIS-empowered networks. 

Moreover, our results can be applied to the design and optimization of new emerging techniques. In the 5G 3GPP New Radio (NR) standard, the optimal transmit beam is determined by a process that includes beam-sweeping and beam-refinement, using a pre-defined set of $m$ analog beams that form the analog codebook \cite{Giordani19}. 
The angular distribution of the users, as per \textbf{Corollary \ref{cor:3D_joint_angular_PDF}}, can be exploited to design the optimal set of $m$ beams. Finally, as another example, our results can be applied to the design of wireless routing, where the marginal distribution of azimuth angles of \textbf{Corollary \ref{cor:marginal_azimuth_PDF}} can be used. In essence, wireless routing aims to transmit a message between two nodes $A$ and $B$ in a wireless multi-hop network \cite{Srinivasa10}. The origin and destination nodes cannot communicate directly due to the limited transmit power that establishes a maximum communicating range, $r_{\rm max}$. The problem is to find the optimal path that minimizes the number of hops. In this scenario, for each node and its given location $u$, the marginal PDF of azimuth angles can be used to find the optimal transmit direction towards the next node.

{As an application example, we will derive the coverage probability, i.e., the CCDF of the SNR for an scenario where the AP is placed at a reference location $u$, and the UEs are randomly placed within a rectangle. The discussions and usefulness of these results are further explored in the numerical results section.}  

\subsection{\textcolor{black}{Coverage probability analysis of directional beamforming in a noise-limited scenario}}
{Let us consider a 2D case, where the AP, placed at $u=(u_x, u_y)$, has a directional antenna gain, $g(\theta)$, whereas the UEs are equipped with omnidirectional antennas. The physical antenna or antenna array has an orientation defined by the bearing angle $\xi$, which points to the direction of maximal gain. With this considerations the SNR of a randomly chosen UE can be expressed as}

\textcolor{black}{
\begin{equation}
    \mathrm{SNR} = \frac{g_t(\Theta - \xi) (\tau R)^{-\alpha} \lvert \beta \lvert^2 \rho_t}{N_0},
\end{equation}}
{\noindent where the fast fading is assumed to follow Rayleigh distribuition, i.e., $\lvert \beta \lvert^2 \sim \mathrm{Exp}(1)$. 
Without loss of generality, we consider the antenna gain expressed in decibels in {\cite{38.901}} as}
\textcolor{black}{\begin{equation}
    \label{eq:antenna_gain}
    g_t(\theta) = g_{\mathrm{max}} - \min \left( \min \left(12 \left(\frac{\theta}{\theta_{\mathrm{3dB}}} \right)^2, a_{\mathrm{max}} \right) \right),
\end{equation}}
{\noindent where $\theta_{\mathrm{3dB}}$ is the half power beamwidth (HPBW) in radians, $g_{\mathrm{max}}$ the maximum directional gain, and $a_{\mathrm{max}}$ the side-lobe attenuation factor.

The coverage probability, $\bar{F}_\mathrm{SNR}(t)$, is given in the following corollary.}

\begin{corollary}
    \label{cor:CCDF_SNR}
    {The coverage probability of a UE randomly placed within a rectangle $\mathfrak{R}(o)$, with omnidirectional antenna pattern, that receives transmission from an AP placed at $u$, with a directional antenna gain $g_t(\theta)$ and bearing angle $\xi$, is given by}:
    \textcolor{black}{\begin{align}
        \bar{F}_\mathrm{SNR}(t) &= \frac{1}{\ell_x \ell_y} \sum_{i=1}^{8} \Bigg( \int_{r=0}^{h_i} r s_i^{(<)}(r) \dd r 
        + \int_{r=h_i}^{r_\mathrm{max}} r s_i^{(\geq)}(r) \dd r \Bigg), \nonumber
    \end{align}}
    \noindent {where $r_\mathrm{max}$ represents the distance between $u$ and the farthest vertex of $\mathfrak{R}(o)$, and}
    \textcolor{black}{\begin{equation}
        s_i^{(\mathrm{lb})}(r) = \mathbbm{1}\Bigg(\chi^{(\mathrm{lb})}_{i,1}(r) < \chi^{(\mathrm{lb})}_{i,2}(r)\Bigg) \int_{r=\chi^{(\mathrm{lb})}_{i,1}(r)}^{\chi^{(\mathrm{lb})}_{i,2}(r)} \mathrm{e}^{\left(-\frac{t N_0 (\tau r)^\alpha}{\rho_t g(\theta - \xi)}\right)} \dd \theta, \nonumber
    \end{equation}}
    \noindent {where $\mathrm{lb} = \{<, \geq\}$}.
    \begin{proof}
        {The coverage probability can be expressed as} 
        \textcolor{black}{\begin{align}
            \bar{F}_\mathrm{SNR}(t) = \Pr\left( {\rm{SNR} \ge t} \right) \overset{\mathrm{(a)}}{=} \mathbb{E}_{R, \Theta} \left[\Pr\left( \lvert \beta \lvert^2 \geq \frac{t N_0 (\tau R)^\alpha}{\rho_t g_t(\Theta - \xi)} \right)  \right], \nonumber 
        \end{align}} 
        \noindent {where (a) comes after reordering the expression of the SNR and conditioning over the pair of RVs, $R$ and $\Theta$. Finally, applying the definition of the CCDF of an exponential distribution, expressing the expectation in integral form with the joint PDF of $R$ and $\Theta$ as per \mbox{\textbf{Corollary \ref{cor:joint_pdf_R_Theta}}} and manipulating the resulting expression to avoid discontinuities over the integration completes the proof.} 
    \end{proof}
\end{corollary}

{In the special case where $\ell_y \gg \ell_x$, the CCDF of the SNR can be approximated with the following expression.}
\begin{approximation} {The CCDF of the SNR when $\ell_y \gg \ell_x$ and the reference node, $u$ placed at the center of the rectangle can be written as follows:}
\label{appr:CCDF_SNR}
\begin{align}
    \bar{F}_\mathrm{SNR}(t) &= \frac{\left(\Gamma(\alpha^{-1}) - \Gamma\left(\alpha^{-1}, 2^{-\alpha}\lambda(\pi/2) (\ell_y \tau)^{\alpha}\right) \right)}
    {\ell_y \alpha \tau \; \sqrt[\alpha]{\lambda(\pi/2)}} \nonumber \\
    &\frac{\left(\Gamma(\alpha^{-1}) - \Gamma\left(\alpha^{-1}, 2^{-\alpha}\lambda(3\pi/2) (\ell_y \tau)^{\alpha}\right) \right)}
    {\ell_y \alpha \tau \; \sqrt[\alpha]{\lambda(3\pi/2)}},
\end{align}
\noindent {where $\Gamma(a,z) = \int_{z}^{\infty}t^{a-1} \mathrm{e}^{-t} \dd{t}$ is the upper incomplete gamma function and $\lambda(\phi) = \frac{t N_0}{\rho_t g(\phi - \xi)}$.}
\begin{proof}
    {Applying \mbox{\textbf{Approximation \ref{appr:joint PDF}}} to the expression given in \mbox{\textbf{Corollary \ref{cor:CCDF_SNR}}} and solving the integrals completes the proof}.
\end{proof}
\end{approximation}

\section{Numerical Results}
\label{sec:Results}

\subsection{Validation}

\begin{table}[t]
\renewcommand{\arraystretch}{1.3}
\caption{Main Parameters}
\label{tab:Sim_scenarios}
\centering
\begin{tabular}{ c c c c }
\toprule
Scenario & Shape ($\ell_x, \ell_y$) & Reference location $u$ & Height $v_z$ \\
\toprule
\textit{O} &  $\ell_x=200, \ell_y=100$ m & $u=(30, 25, 10)$ m & $v_z=1.5$ m \\
\hline
\textit{A} &  $\ell_x=200, \ell_y=9.75$ m & $u=(0, 4.8)$ & - \\
\hline
\textit{B} &  $\ell_x=3, \ell_y=5$ m & $u=(0.5, 1.25, 3)$ m & $v_z=1.5$ m \\
\hline
\textit{C} &  $\ell_x=200, \ell_y=100$ m & $u=(30, 25, 10)$ m & $v_z=120$ m \\
\bottomrule
\end{tabular}
\end{table}
%

We now evaluate the theoretical expressions previously derived for the cases of four exemplary regions that model different scenarios as summarized in Table \ref{tab:Sim_scenarios}. 

\begin{figure*}[t]%
    \centering
    \subfloat[]{{\includegraphics[width=0.25\textwidth]{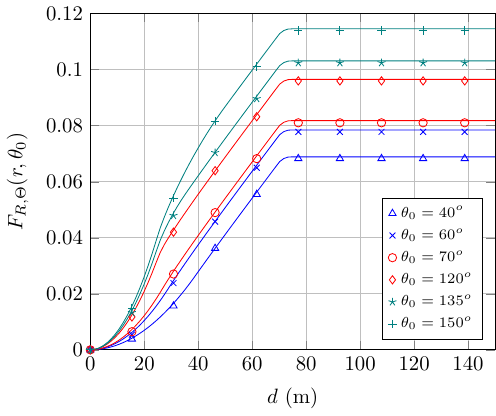} }}%
    \subfloat[]{{\includegraphics[width=0.25\textwidth]{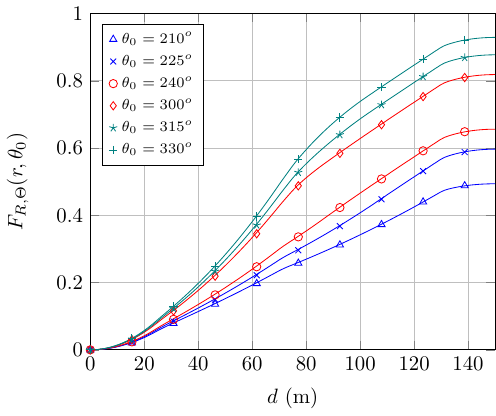} }}%
    \subfloat[]{{\includegraphics[width=0.25\textwidth]{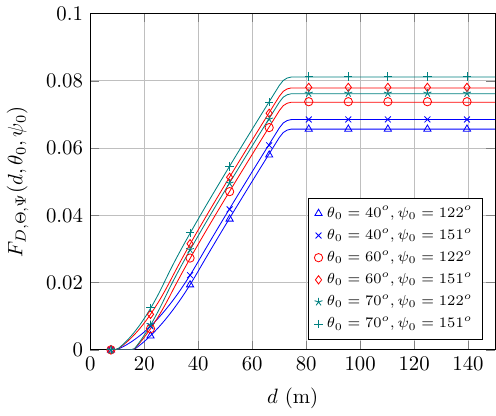} }}%
    \subfloat[]{{\includegraphics[width=0.25\textwidth]{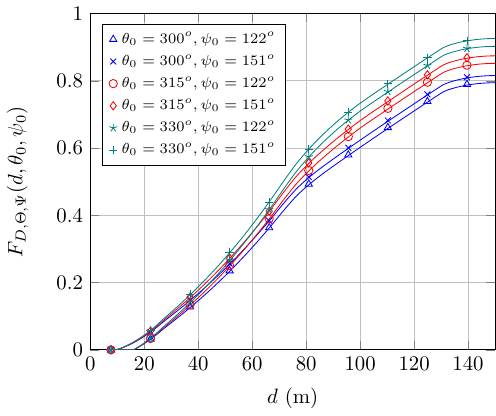} }}%
    \caption{Joint CDF of distance in the $xy$ plane, $R$, and azimuth angle, $\Theta$, particularized for angles within the first and second quadrants (a), and third and fourth quadrants (b); and joint CDF of distance, $D$, azimuth and zenith angles, $\Theta, \Psi$, particularized for azimuth angles within the first (c), and fourth (d) quadrants.}%
    \label{fig:validation_all}%
\end{figure*}

Throughout this section, theoretical results are validated and double-checked with Monte Carlo (MC) simulations\footnote{The code is available at: \url{https://github.com/franmarve/joint_pdf_dist_angles}}. The empirical distributions have been estimated using $10^5$ realizations of random points. We first consider a 2D scenario (\textit{O} in Table \ref{tab:Sim_scenarios}), for which we represent the joint CDF of the distance on the $xy$ plane, $R$, and the azimuth angle, $\Theta$, given by \textbf{Theorem \ref{theor:Joint CDF}}. The shape of this scenario matches a typical public square in many cities, (e.g., the Dam Square in Amsterdam). We particularize the expression for a set of three angles per quadrant for the variable $\theta=\theta_0$ (see Fig. \ref{fig:validation_all} (a) and (b)). 
For the 3D case, the joint CDF of distance, $D$, azimuth and zenith angles, $\Theta, \Psi$ (\textbf{Theorem \ref{theo:Joint_CDF_3D}}) is validated in Fig. \ref{fig:validation_all} (c) and (d). 
We observe an excellent match between simulation and theoretical results.

\begin{figure*}[t]%
    \centering
    \subfloat[]{{\includegraphics[width=0.3\textwidth]{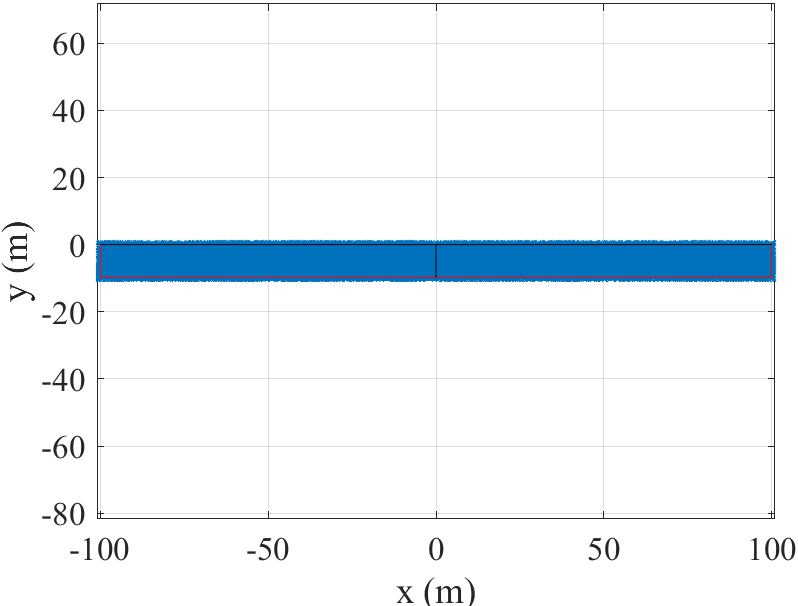} }}%
    \subfloat[]{{\includegraphics[width=0.3\textwidth]{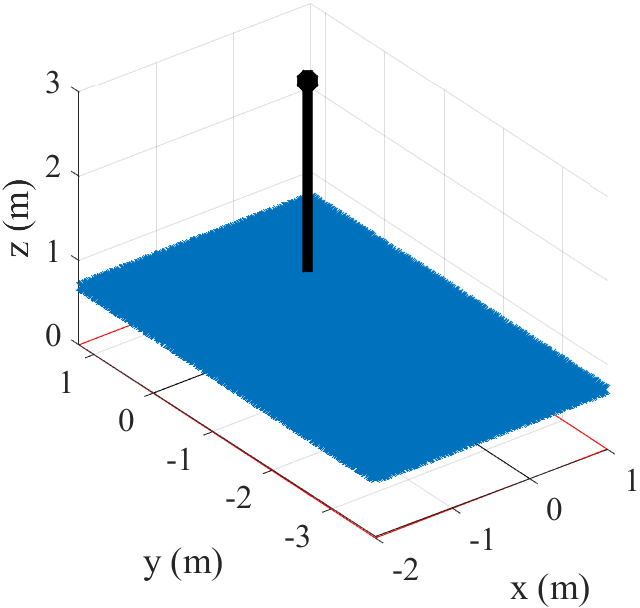} }}%
    \subfloat[]{{\includegraphics[width=0.3\textwidth]{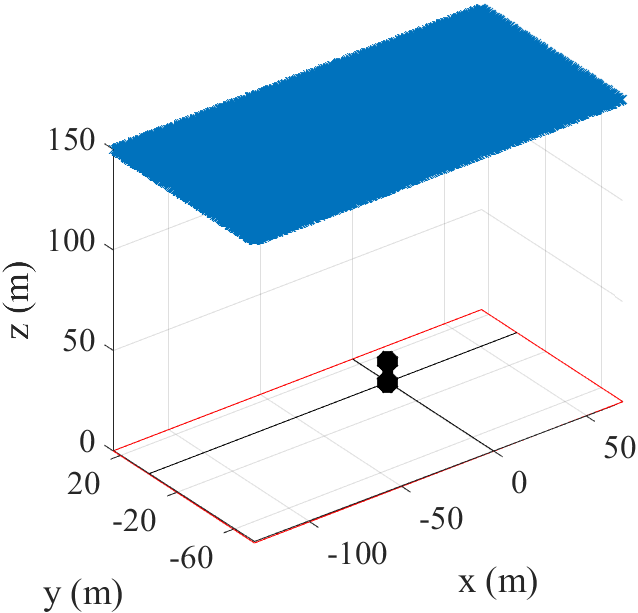} }}%
    \caption{Sketch of reference scenarios A, B, C as per Table \ref{tab:Sim_scenarios}: (a) 2D scenario which models a road with $3$ lanes (each lane is $3.25$ m wide) and a road side unit located close to the edge of the road; (b) indoor office scenario, with a room of $3 \times 5$ m, and an AP placed on the ceiling at the height of $3$ m; (c) UAV-based network, with drones flying at the height of $120$ m, that communicate with a BS antenna ($10$ m high). The reference node is drawn in black, whereas the random nodes are drawn in black. All the axes have the same scale.}%
    \label{fig:scenarios2}%
\end{figure*}



\subsection{Reference Scenarios}

\begin{figure*}[t]%
    \centering
    \subfloat[]{{\includegraphics[width=0.3\textwidth]{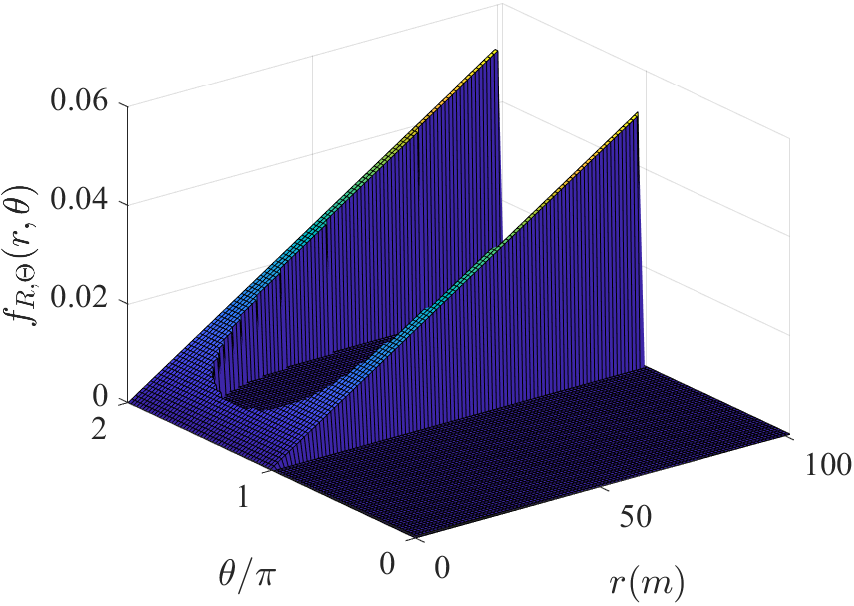} }}%
    \subfloat[]{{\includegraphics[width=0.3\textwidth]{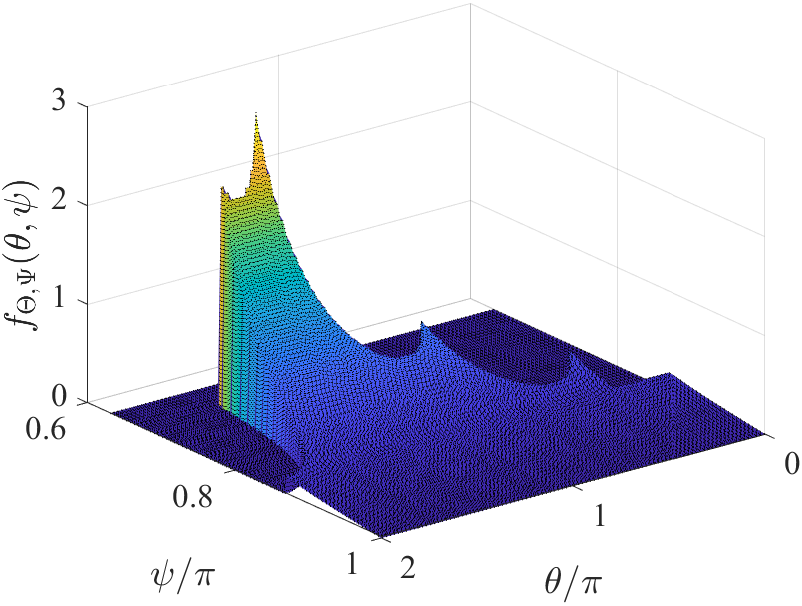} }}%
    \subfloat[]{{\includegraphics[width=0.3\textwidth]{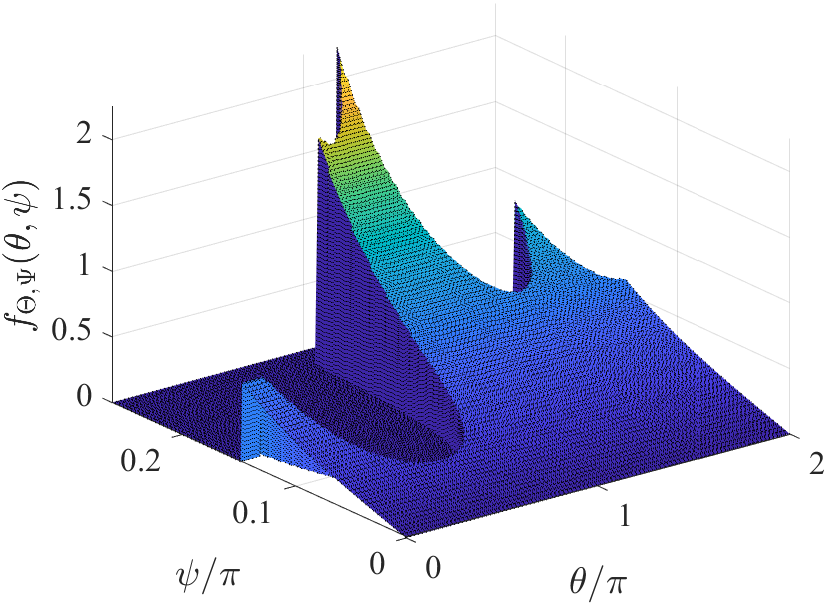} }}%
    \caption{Joint PDF for different scenarios: (a) joint PDF of distance on the $xy$ plane, $R$, and azimuth angle, $\Theta$, for scenario \textit{A}; (b) and (c) joint PDF of azimuth and zenith angles, $\Theta, \Psi$, for scenarios \textit{B} and \textit{C} respectively.}%
    \label{fig:scenarios_results_all}%
\end{figure*}

After validating the 2D and 3D distributions, we now turn our attention to the scenarios defined in Table \ref{tab:Sim_scenarios}. Fig. \ref{fig:scenarios2} illustrates scenarios \textit{A}, \textit{B} and \textit{C}. Scenario \textit{A} represents a road segment of $200$ m with $3$ lanes of $3.25$ m each, and an RSU on the edge of the road, as shown in Fig. \ref{fig:scenarios2}-(a).
The joint PDF of distance and azimuth angle, given in \textbf{Corollary \ref{cor:joint_pdf_R_Theta}}, is numerically evaluated and illustrated in Fig. \ref{fig:scenarios_results_all}-(a). 
We can see how the joint distribution depends on the shape of the region and the location of the arbitrary node. A high correlation between the distance and azimuth angle is observed. Longer distances, $R$, are associated with azimuth angles close to either $\pi$ or $2\pi$ radians, since those directions point to the two extremes of the road segment. The minimum distance values ($R$) are associated with azimuth angles around $3\pi/2$ radians, since in this direction the reference node points to the perpendicular of the road segment, which has a width of only \mbox{$9.75$ m}. It is also observed that the joint PDF is null for azimuth angles between $0$ and $\pi$ since those directions point outside the road, and thus they are forbidden locations for the vehicles (i.e., random nodes). This joint PDF identifies the direction (or azimuth angles) where a higher user density is expected, and this can be used in the design of beam patterns. 



The joint PDF of azimuth, $\Theta$, and zenith, $\Psi$, angles, given by \textbf{Corollary \ref{cor:3D_joint_angular_PDF}}, is illustrated in Fig. \ref{fig:scenarios_results_all}-(b) for scenario \textit{B}. This scenario, shown in Fig. \ref{fig:scenarios2}-(b), models an office room of $3 \times 5$ m, with an AP placed on the ceiling at a height of $3$ m. The geometry of this scenario sets a range of zenith angles $\Psi \in (0.63 \pi, \pi)$, as observed in Fig. \ref{fig:scenarios_results_all}-(b). The highest node density is obtained for $\theta=1.34\pi$ and $\psi=0.66\pi$, representing the direction where users are most likely to be found. This information can be exploited for the design of beam patterns and wireless routing as discussed in section \ref{sec:applications}. 



Lastly, scenario \textit{C} (cfr. Fig. \ref{fig:scenarios2}-(c)) represents a UAV-based network, with drones flying at a height of $120$ m, that communicate with a BS antenna ($10$ m high). Here, the random nodes are higher (in altitude) than the reference node (BS antenna), resulting in zenith angles within the range $\Psi\in (0,\psi_{\max})$, with $\psi_{\max}=0.26\pi$. This is observed in Fig. \ref{fig:scenarios_results_all}-(c), which represents the joint PDF of azimuth and zenith angles. In this scenario, the direction of maximum node density is given by $\theta=1.16 \pi, \psi = 0.26 \pi$. On the other hand, minimal user density is observed at azimuth angles around $\theta=\pi/2$. This is due to the fact that the reference node is close to the edge of the region (on the positive $y$ axis direction) where the drones are. 


\subsection{Impact of shape and location on the distribution}

\begin{figure}[t]%
    \centering
    \subfloat[]{{\includegraphics[width=0.5\columnwidth]{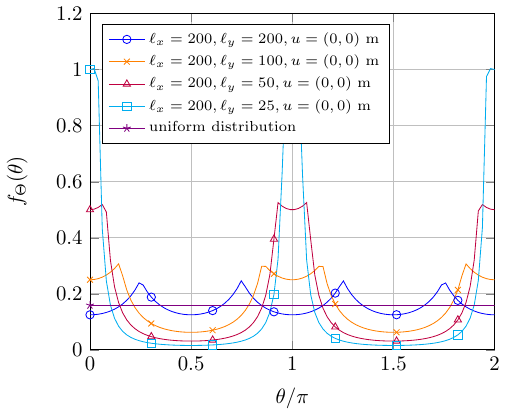} }}%
    \subfloat[]{{\includegraphics[width=0.5\columnwidth]{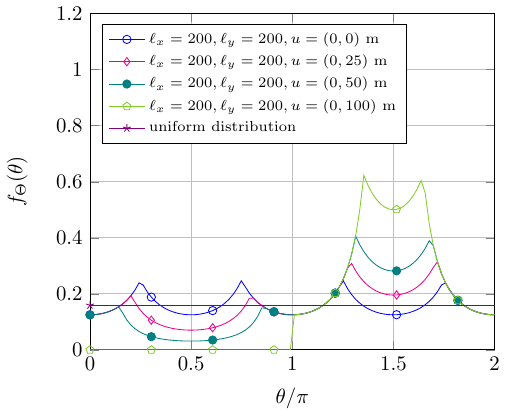} }}%
    \caption{Marginal PDF of $\theta$ to illustrate the effect of the shape of the region (a), and the effect of the location of the arbitrary point $u$ (reference node) (b).}%
    \label{fig:3_f_Theta_all}%
\end{figure}


Fig. \ref{fig:3_f_Theta_all}-(a) shows the marginal PDF of the azimuth angle, given in \textbf{Corollary \ref{cor:marginal_azimuth_CDF}}, for a reference node located at the region's center of mass. To study the effect of the region's shape, the PDF is evaluated for different lengths of the $y$-axis side, $\ell_y$, while the other side ($\ell_x$) remains constant. The PDF of a uniform distribution, typically assumed in the
literature, is also included for comparison. 

We first observe that the uniform assumption is not realistic { even if the shape is perfectly regular (i.e., a square $\ell_x=\ell_y$). In this case, the part of the square area exceeding the inscribed circle (close to the vertices and representing $ \left(1- \pi/4\right) $ times the total square area) adds peaks to the angle distribution in $\pi/4$ and its odd multiples, whichever the side length. 
The distribution deviations from the uniform distribution are more pronounced as the shape of the region becomes more irregular, i.e., as one of the sides becomes greater than the other one. As the $\ell_y$ side is reduced the node density increases in the positive and negative directions of the $x$ axis, which leads to a clear concentration of nodes at azimuth angles close to $\theta=0$, $\theta=2\pi$ and $\theta=\pi$}. 

Finally, the effect of the node location is investigated in Fig. \ref{fig:3_f_Theta_all}-(b) for {a} square region ($200 \times 200$ m). Initially, a reference node placed at the center of mass is considered, and then its position is modified to approach the edge of the region in the positive $y$ axis direction. Again, the uniform distribution{, i.e., a circular shape with the fixed point at its center,} is included for comparison. 
The distribution becomes more different from the uniform one as the reference node moves away from the center of mass. Additionally, as the reference node approaches the edge in the positive $y$ axis direction, the (random) node density increases in the opposite direction, i.e., $\theta=3\pi/2$, while it reduces in the direction of movement, i.e., $\theta=\pi/2$. In the extreme case where the reference node is placed at the edge of the region, i.e., $u=(0,100)$, the marginal PDF is null for angles in the range $(0,\pi)$ since those directions point outside the region. 

\subsection{\textcolor{black}{Directional beamforming in finite wireless networks}}
\label{sec:Results_CCDF_SNR}

\begin{figure*}[t]%
    \centering
    \subfloat[]{{\includegraphics[width=0.3\textwidth]{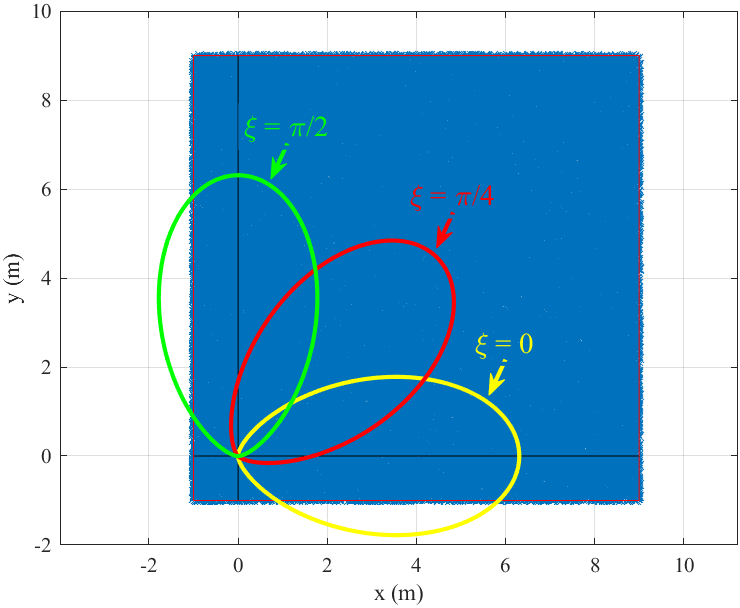} }}%
    \subfloat[]{{\includegraphics[width=0.3\textwidth]{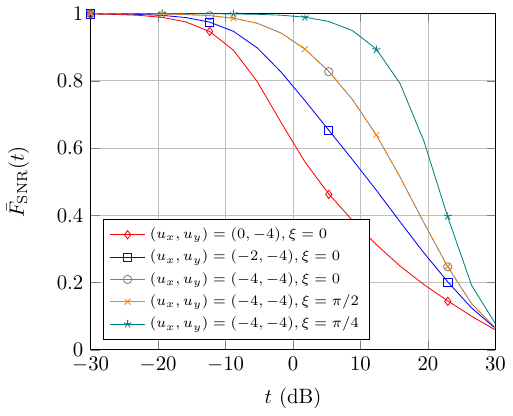} }}%
    \subfloat[]{{\includegraphics[width=0.3\textwidth]{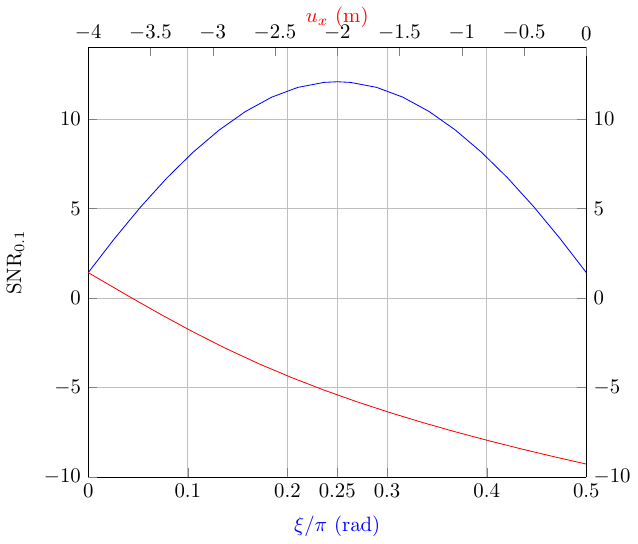} }}%
    \caption{{Results related to the analysis of directional beamforming: (a) graphical representation of a finite area network with $(l_x,l_y) = (10,10)$ m, $u=(-4,-4)$ for 3 different bearing angles of the directive antenna ($\xi=\{0, \pi/4, \pi/2\}$); (b) theoretical (thin and dashed lines) and simulation results (markers) of the CCDF of the SNR (i.e., coverage probability) for different AP locations and bearing angles; (c) $10\%$-percentile of the SNR versus the location of the AP (red line: $u_x \in [-4, 0]$, $u_y=-4$, $\xi=0$), and the bearing angle (black line: $\xi \in [0, \pi/2]$, $u_x=-4$, $u_y=-4$)}}%
    \label{fig:CCDF_SNR_figures}%
\end{figure*}

\begin{figure}[t]
    \centering
    \includegraphics[width=0.99\columnwidth]{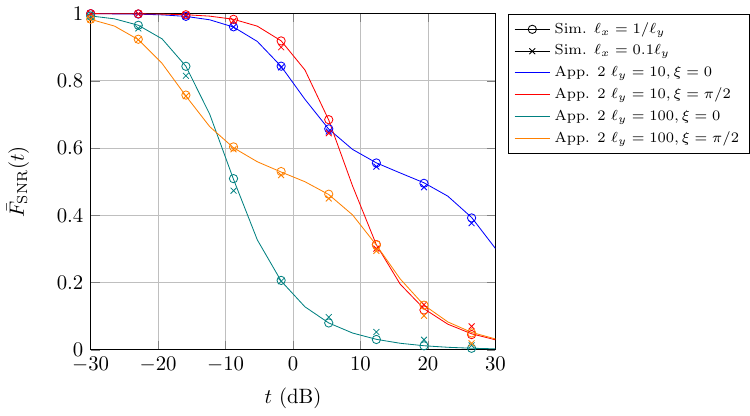}
    \caption{{CCDF of the SNR when $\ell_y \gg \ell_x$ as per \mbox{\textbf{Approximation \ref{appr:CCDF_SNR}}} (thin line) for $\ell_y=100$ and $\ell_y=10$ with bearing angles of $\xi = \{0, \pi/2\}$. Simulation results are shown with markers for $\ell_x = 1/\ell_y$ (circles) and $\ell_x = 0.1 \ell_y$ (crosses).}}
    \label{fig:CCDF_SNR_appr}
\end{figure}

{In this subsection we investigate the impact of the AP location, $u$, and bearing angle, $\xi$, on the coverage probability, i.e., the CCDF of the SNR, given in \mbox{\textbf{Corollary \ref{cor:CCDF_SNR}}}. In addition, we highlight the usefulness of the obtained results for finite wireless networks design. 
For this numerical application example we consider an indoor office room, with highly isolated walls (i.e., noise limited communication system) of side length $10$ m. We aim at determining the optimal AP location, $u$, and bearing angle of the AP's antenna, $\xi$, that maximize the SNR for $10\%$ of the worst UEs. Thus, this metric is mathematically expressed as the $10\%$-percentile of the SNR, which can be written as} 
\textcolor{black}{\begin{equation}
    \mathrm{SNR}_{p} = \{ t \in \mathbb{R} \lvert 1 - \bar{F}_\mathrm{SNR}(t) - p = 0 \},
\end{equation}}
where $p=0.1$. 
For this exercise, we assume that there are some restrictions on the valid AP locations due to implementation issues. More specifically, in this example we assume that the AP must lie at $1$ m of the right hand side wall of Fig. \mbox{\ref{fig:CCDF_SNR_figures}}-(a). This involves that the valid locations are expressed as $u=(u_x,-4)$. Nevertheless, there are no restrictions on the bearing angles in this example, i.e., $\xi \in [-\pi, \pi]$. 
Regarding the path loss and antenna gain, we consider the 3GPP indoor office hotspot model as per \mbox{\cite{38.901}} with a carrier frequency of $2$ GHz, a maximal gain of $g_\mathrm{max}=8$ dBi, an HPBW of $\theta_\mathrm{3dB}=65$ degrees, and a side-lobe attenuation factor of $a_\mathrm{max}=-22$ dB.

{Fig. \mbox{\ref{fig:CCDF_SNR_figures}}-(a) illustrates the considered scenario for 3 different bearing angles. The antenna gain, given by \mbox{\eqref{eq:antenna_gain}}, is shown with different colors for $\xi=\{0, \pi/4, \pi/2\}$. The CCDF of the SINR is illustrated in Fig. \mbox{\ref{fig:CCDF_SNR_figures}}-(b). A perfect match between simulation and theoretical results is observed. Besides, we can see that AP locations and bearing angles pointing to a greater UE density lead to a higher coverage probability. Finally, the applicability of the obtained expressions to determine the optimal AP and bearing angle is shown in Fig. \mbox{\ref{fig:CCDF_SNR_figures}}-(c). 
The highest value is obtained for $u=(-4, -4), \xi = \pi/4$. Remark that the difference in $\mathrm{SNR}_{0.1}$ between the worst configuration ($-9.27$ dB with $u_x=0, \xi=0$ or $\xi=\pi/2$) and optimal configuration ($12.08$ dB with $u_x=-4, \xi=\pi/4$) is $21.35$ dB, which highlights the importance of a proper selection of the AP location and bearing angle.}

{Finally, we assess the accuracy of the expression given by \mbox{\textbf{Approximation \ref{appr:CCDF_SNR}}} for the case when $\ell_y \gg \ell_x$, for different side lengths $\ell_y = \{10, 100\}$ and bearing angles $\xi = \{0, \pi/2\}$. The results are illustrated in Fig. \mbox{\ref{fig:CCDF_SNR_appr}} which shows a good match between analytical and simulation results for lengths of the horizontal axis, $\ell_x$ up to $10\%$ of the vertical axis, $\ell_y$, while the match between analytical and simulation is perfect when $\ell_x=1/\ell_y$.}




\section{Conclusion}
\label{sec:Conclusions}
Considering finite wireless networks, we have proven a non-negligible correlation between the distance and azimuth angle that characterize the link between an arbitrarily placed reference node and randomly located nodes. 
We have first proposed a mathematical framework for analyzing the joint distribution 
in 2D arbitrarily-shaped regions. We then particularized this framework for the relevant case of a rectangular region. 
We further extended our results to consider the 3D case where the zenith angle must be considered jointly with the distance and azimuth angle. To illustrate the importance of the proposed framework, a number of relevant applications have been identified. 
We have presented some numerical results to validate the theoretical expressions and shed light on the dependencies between the distance and angles {as well as} on the effect of the {region's} shape and {the} location of the reference node. {Finally, we have illustrated the usefulness of the results in finite wireless networks by analyzing the coverage probability of an indoor scenario and determining the optimal AP location and antenna bearing angle.} 
%

\appendices

\section{Proof of Lemma \ref{lem:Rectangle_in_polar}}
\label{App:Proof of lem:Rectangle_in_polar}
        The rectangular region can be expressed in Cartesian coordinates as follows
        \begin{align}
            \mathfrak{R}(-u) = \Bigg\{(x,y) \in \mathbb{R}^2 &\Big\lvert -\frac{\ell_x}{2} - u_x \leq x \leq \frac{\ell_x}{2} - u_x \nonumber \\
                &\wedge -\frac{\ell_y}{2} - u_y \leq y \leq \frac{\ell_y}{2} - u_y \Bigg\}.
        \end{align}
        %
        %
        This set can be written in polar coordinates using \eqref{eq:polar_coord} in the form of \eqref{eq:Convex_region_in_polar}, where the condition $\mathcal{B}(\rho,\phi)$ is expressed as
        \begin{align}
            \label{eq:4_ineq_polar}
             \mathcal{B}(\rho,\phi) &= -\frac{\ell_x}{2} - u_x \leq \rho \cos(\phi) \leq \frac{\ell_x}{2} - u_x \nonumber \\
                &\wedge -\frac{\ell_y}{2} - u_y \leq \rho \sin(\phi) \leq \frac{\ell_y}{2} - u_y.
        \end{align}
        Then, we isolate the variable $\rho$ from $\mathcal{B}(\rho, \phi)$ as follows
        \begin{align}
            \Bigg[\underbrace{\Big(\cos(\phi) \geq 0\Big)}_{\mathcal{A}_c(\phi)} \wedge \underbrace{\left(\rho \leq \frac{h_{x^+}}{\cos(\phi)} \right)}_{\mathcal{A}_{x^+}(\rho,\phi)} 
            \vee \bar{\mathcal{A}_c}(\phi) \wedge \underbrace{\left(\rho \leq \frac{h_{x^-}}{\cos(\phi)} \right)}_{\mathcal{A}_{x^-}(\rho,\phi)}\Bigg]
            \nonumber \\
            \wedge \Bigg[\underbrace{\Big(\sin(\phi) \geq 0\Big)}_{\mathcal{A}_s(\phi)} \wedge \underbrace{\left(\rho \leq \frac{h_{y^+}}{\sin(\phi)} \right)}_{\mathcal{A}_{y^+}(\rho,\phi)} 
            \vee \bar{\mathcal{A}_s}(\phi) \wedge \underbrace{\left(\rho \leq \frac{h_{y^-}}{\sin(\phi)} \right)}_{\mathcal{A}_{y^-}(\rho,\phi)} \Bigg].
        \end{align}
        \noindent after some manipulations over the inequalities in \eqref{eq:4_ineq_polar}. It should be noticed that each of the former inequalities yields two inequalities but one of them is always true, and thus it is discarded. 
        Then, applying the distributive and associative properties of the logical operators, it leads to
        \begin{align}
            \mathcal{B}(\rho, \phi) &= \underbrace{\mathcal{A}_c(\phi) \wedge \mathcal{A}_s(\phi)}_{=\phi \in \mathfrak{Q}_1} \wedge \mathcal{A}_{x^+}(\rho,\phi) \wedge \mathcal{A}_{y^+}(\rho,\phi) \nonumber \\
            &\vee \left(\phi \in \mathfrak{Q}_4\right) \wedge \mathcal{A}_{x^+}(\rho,\phi) 
                \wedge \mathcal{A}_{y^-}(\rho,\phi) \nonumber \\
            &\vee \left(\phi \in \mathfrak{Q}_2\right)  \wedge \mathcal{A}_{x^-}(\rho,\phi) 
                \wedge \mathcal{A}_{y^+}(\rho,\phi) \nonumber \\
            &\vee  \left( \phi \in \mathfrak{Q}_3 \right)  \wedge \mathcal{A}_{x^-}(\rho,\phi) 
                \wedge \mathcal{A}_{y^-}(\rho,\phi).
        \end{align}
        %
        %
        It is identified that the \texttt{and} operation of the logical expressions $\mathcal{A}_c(\phi)$ and $\mathcal{A}_s(\phi)$ and its negations are equivalent to checking whether the angle $\phi$ belongs to each of the angular quadrants. For instance, the first term can be expressed as follows: $\mathcal{A}_c(\phi) \wedge \mathcal{A}_s(\phi) = \phi \in \mathfrak{Q}_1$. The remaining terms are manipulated as follows
        \begin{align}
            \label{eq:A_x}
            \mathcal{A}_{x^+}(\rho,\phi) 
            \wedge \mathcal{A}_{y^+}(\rho,\phi) &= \rho \leq \frac{h_{x^+}}{\cos(\phi)} \wedge \rho \leq \frac{h_{y^+}}{\sin(\phi)}  \nonumber \\
            &= \rho \leq \min \left(\frac{h_{x^+}}{\cos(\phi)}, \frac{h_{y^+}}{\sin(\phi)} \right).
        \end{align}
        \noindent 
        The terms $\mathcal{A}_{x^-}(\rho,\phi) \wedge \mathcal{A}_{y^+}(\rho,\phi)$, 
			$\mathcal{A}_{x^+}(\rho,\phi) \wedge \mathcal{A}_{y^-}(\rho,\phi)$ and $\mathcal{A}_{x^-}(\rho,\phi) \wedge \mathcal{A}_{y^-}(\rho,\phi)$ lead to similar expressions to \eqref{eq:A_x} but using different combinations of the parameters $h_{x^+}$, $h_{y^+}$, $h_{x^-}$, and $h_{y^-}$. Details are omitted for the sake of compactness. Leveraging this fact and substituting the functions $h_{x}(\phi)$ and $h_{y}(\phi)$ in the above expression completes the proof.  

\section{Proof of Lemma \ref{lem:Step2}}
\label{App:Proof of lem:Step2}
        The boolean expression $\mathcal{B}(r,\phi) = r \leq \beta(\phi)$ can be expressed as
        \begin{align}
            \label{eq:proof_B}
            \mathcal{B}(r,\phi) &\overset{\mathrm{(a)}}{=} r \leq \Bigg( \frac{h_x(\phi)}{\cos(\phi)} \mathbbm{1} \Big( \mathcal{C}(\phi) \Big)  + \frac{h_y(\phi)}{\sin(\phi)} \mathbbm{1} \Big( \bar{\mathcal{C}}(\phi) \Big) \Bigg)  \\
            &\overset{\mathrm{(b)}}{=} \underbrace{ \left(\frac{h_x(\phi)}{\cos(\phi)} \geq r\right)}_{\mathcal{D}(\phi)}  \wedge \mathcal{C}(\phi) \vee 
            \underbrace{\left( \frac{h_y(\phi)}{\sin(\phi)} \leq r \right)}_{\mathcal{E}(\phi)} \wedge \bar{\mathcal{C}}(\phi), \nonumber
        \end{align}
        \noindent where (a) comes after expressing the $\min(x,y)$ function as the sum of two indicator functions with $\mathcal{C}(\phi) =  \frac{h_x(\phi)}{\cos(\phi)} \leq \frac{h_y(\phi)}{\sin(\phi)}$, and (b) after reordering and applying the inequality in both terms. 
        Below, we isolate $\phi$ in the boolean expressions $ \mathcal{C}(\phi),  \mathcal{D}(\phi),  \mathcal{E}(\phi)$. 
        The term $\mathcal{C}(\phi)$ can be written as
        \begin{align}
        \label{eq:C_phi}
            \mathcal{C}(\phi) &= 
                \overbrace{\underbrace{\left( \tan(\phi) \leq \frac{h_y(\phi)}{h_x(\phi)} \right)}_{\mathcal{F}_1(\phi)} \wedge \mathcal{A}_t(\phi)}^{\mathcal{C}_1(\phi)} \nonumber \\
                &\vee \overbrace{\underbrace{\left( \tan(\phi) \geq \frac{h_y(\phi)}{h_x(\phi)} \right)}_{\mathcal{F}_2(\phi)} \wedge \bar{\mathcal{A}}_t(\phi)}^{\mathcal{C}_2(\phi)},
        \end{align}
        \noindent where the expression of $\mathcal{C}(\phi)$ has been manipulated to group the $\sin(x)$ and $\cos(x)$ as a tangent function, and this latter term is isolated by splitting the inequality into the cases where the tangent is positive ($\mathcal{A}_t(\phi) = \mathtt{1}$) and negative ($\bar{\mathcal{A}_t}(\phi) = \mathtt{1}$). Thus, $\mathcal{A}_t(\phi) = \phi \in \mathfrak{Q}_1 \cup \mathfrak{Q}_3$ and $\bar{\mathcal{A}_t}(\phi) =  \phi \in \mathfrak{Q}_2 \cup \mathfrak{Q}_4$.
        The solution of the inequality $\mathcal{F}_1(\phi)$ is $\phi \in \left\{\left(-\frac{\pi}{2} + \pi n, \atan(\frac{h_y(\phi)}{h_x(\phi)}) + n\pi  \right] , n\in \mathbb{Z} \right\}$, whereas the solution of $\mathcal{F}_2(\phi)$ is $\phi \in \left\{ \left[\atan(\frac{h_y(\phi)}{h_x(\phi)}) + n\pi, \frac{\pi}{2} + n \pi\right), n\in  \mathbb{Z} \right\}$. Thus, the intersection of these solutions with the intervals that define $\mathcal{A}_t(\phi)$ and $\bar{\mathcal{A}_t}(\phi)$ leads to
        \begin{align}
            \label{eq:C1 and C2}
            \mathcal{C}_1(\phi) &= \phi \in \left[0, \atan(\frac{h_y(\phi)}{h_x(\phi)})\right) \cup \left(\pi, \atan(\frac{h_y(\phi)}{h_x(\phi)}) \right],  \\
            \mathcal{C}_2(\phi) &= \phi \in \left[\atan(\frac{h_y(\phi)}{h_x(\phi)}) + \pi, \pi \right) \cup \left[\atan(\frac{h_y(\phi)}{h_x(\phi)}) + 2\pi, 2\pi \right). \nonumber
        \end{align}
        Similarly, $\bar{\mathcal{C}}(\phi)$ can be expressed as follows
        \begin{align}
            \label{eq:bar_C}
            \bar{\mathcal{C}}(\phi) = \frac{h_x(\phi)}{\cos(\phi)} > \frac{h_y(\phi)}{\sin(\phi)} = \mathcal{C}_3(\phi) \vee \mathcal{C}_4(\phi).
        \end{align}
        %
        %
        We now isolate $\phi$ on $\mathcal{D}(r, \phi)$ as follows
        \begin{align}
            \label{eq:D}
            \mathcal{D}(r, \phi) = \underbrace{\left(\cos(\phi) \leq \tfrac{h_{x^+}}{r} \right) \wedge \mathcal{A}_c(\phi)}_{\mathcal{D}_1(\phi)} \vee 
            \underbrace{\left(\cos(\phi) \geq \tfrac{h_{x-}}{r} \right) \wedge \bar{\mathcal{A}_c}(\phi)}_{\mathcal{D}_2(\phi)}.
        \end{align}
        Solving the two inequalities leads to
        \begin{align}
            \label{eq:D1 and D2}
            \mathcal{D}_1(r, \phi) &= (r < h_{x^+}) \wedge \mathcal{A}_c(\phi) \vee (r \geq h_{x^+}) 
                \nonumber \\ &\wedge \phi \in \left[\acos(\frac{h_{x^+}}{r}), \frac{\pi}{2} \right] \cup \left[\frac{3\pi}{2}, 2\pi - \acos(\frac{h_{x^+}}{r}) \right], \nonumber \\
            \mathcal{D}_2(r, \phi) &= (r < -h_{x^-}) \wedge \bar{\mathcal{A}_c}(\phi) \vee (r \geq -h_{x^-}) 
            \nonumber \\ &\wedge \phi \in \left[\frac{\pi}{2}, \acos(\frac{h_{x^-}}{r}) \right] \cup \left[2\pi - \acos(\frac{h_{x^-}}{r}), \frac{3\pi}{2} \right].
        \end{align}
        Analogously, $\mathcal{E}(r, \phi)$ can be written as
        \begin{align}
            \label{eq:E}
            \mathcal{E}(r, \phi) = \underbrace{\left(\sin(\phi) \leq \tfrac{h_{y^+}}{r} \right) \wedge \mathcal{A}_s(\phi)}_{\mathcal{E}_1(\phi)} \vee 
            \underbrace{\left(\sin(\phi) > \tfrac{h_{y^-}}{r} \right) \wedge \bar{\mathcal{A}_s}(\phi)}_{\mathcal{E}_2(\phi)}.
        \end{align}
        Then, we solve the above inequalities which yields to
        \begin{align}
            \label{eq:E1 and E2}
            \mathcal{E}_1(r, \phi) &= (r < h_{y^+}) \wedge \mathcal{A}_s(\phi) \vee (r \geq h_{y^+}) \nonumber \\
                &\wedge \phi \in \left[0, \acos(\frac{h_{y^+}}{r}) \right] \cup \left[\pi - \asin(\frac{h_{y^+}}{r}), \pi \right], \nonumber \\
            \mathcal{E}_2(r, \phi) &= (r < -h_{y^-}) \wedge \bar{\mathcal{A}_s}(\phi) \vee (r \geq -h_{y^-}) \nonumber \\ 
                &\wedge \phi \in \left[2\pi + \asin(\frac{h_{y^-}}{r}), 2\pi \right] \cup \left[\pi, \pi - \sin(\frac{h_{y^-}}{r}) \right].
        \end{align}   
        Now that we have written the boolean expressions $\mathcal{C}(\phi)$, $\bar{\mathcal{C}}(\phi)$, $\mathcal{D}(r, \phi)$ and $\mathcal{E}(r, \phi)$ as the \texttt{or} of two boolean expressions, we can write $\mathcal{B}(r, \phi)$ as follows
        \begin{align}
            \label{eq:final_proof_B}
            \mathcal{B}(r, \phi) 
            %
            &= \overbrace{\mathcal{D}_1(r, \phi) \wedge \mathcal{C}_1(\phi)}^{\mathcal{X}_1(r, \phi)} \vee 
                \overbrace{\mathcal{D}_1(r, \phi) \wedge \mathcal{C}_2(\phi)}^{\mathcal{X}_2(r, \phi)} \nonumber \\
            &\vee \mathcal{D}_2(r, \phi) \wedge \mathcal{C}_1(\phi) \vee \mathcal{D}_2(r, \phi) \wedge \mathcal{C}_2(\phi) \nonumber \\
            &\vee \mathcal{E}_1(r, \phi) \wedge \mathcal{C}_3(\phi) \vee \mathcal{E}_1(r, \phi) \wedge \mathcal{C}_4(\phi) \nonumber \\
            &\vee \underbrace{\mathcal{E}_2(r, \phi) \wedge \mathcal{C}_3(\phi)}_{\mathcal{X}_7(r, \phi)} \vee 
                \underbrace{\mathcal{E}_2(r, \phi) \wedge \mathcal{C}_4(\phi)}_{\mathcal{X}_8(r, \phi)},
        \end{align}
        \noindent where expressions of $\mathcal{C}(\phi)$, $\bar{\mathcal{C}}(\phi)$, $\mathcal{D}(r, \phi)$ and $\mathcal{E}(r, \phi)$ in \eqref{eq:proof_B} are used, and the distributive and associative properties of the logical operators are applied. Now, we identify in \eqref{eq:final_proof_B} the terms ranging from $\mathcal{X}_1(r, \phi)$ up to $\mathcal{X}_8(r, \phi)$. This terms, $\mathcal{X}_i(r, \phi)\, \forall i \in [1,8] \subset \mathbb{Z}$, are related to the regions $\mathfrak{X}_i(r, \phi)\, \forall i \in [1,8] \subset \mathbb{Z}$ given in \eqref{eq:X_i} and \eqref{eq:chi_i(r)} as follows:
        \begin{equation}
            \mathfrak{X}_i(r) = \{ \phi \in [0,2\pi] \lvert \mathcal{X}_i(r, \phi) = \mathtt{1} \}.
        \end{equation}
        Finally, substituting \eqref{eq:C1 and C2}, 
        \eqref{eq:D1 and D2}, and \eqref{eq:E1 and E2} in \eqref{eq:final_proof_B} leads to the \texttt{and} operation of $8$ terms with the following form:
        \begin{align}
            \mathcal{X}_i(r, \phi) &= (r < h_i) \wedge \phi \in \Big[\chi^{(<)}_{i,1}(r), \chi^{(<)}_{i,2}(r) \Big) \nonumber \\
                &\vee (r \geq h_i) \Big(\chi^{(\geq)}_{i,1}(r), \chi^{(\geq)}_{i,2}(r) \Big).
        \end{align}
        Identifying the resulting terms $h_i$, $\chi^{(<)}_{i,1}(r)$, $\chi^{(<)}_{i,2}(r)$ $\chi^{(\geq)}_{i,1}(r)$, and $\chi^{(\geq)}_{i,2}(r)$ on each of the $8$ expressions completes the proof for $\mathfrak{T}_{\mathfrak{B}(r)}$. 
        Following a similar approach for $\mathfrak{T}_{\bar{\mathfrak{B}(r)}}$, we can write $\bar{\mathcal{B}}(r,\phi)$ as 
        \begin{equation}
            \bar{\mathcal{B}}(r,\phi) = \bar{\mathcal{D}}(r, \phi) \wedge \mathcal{C}(\phi) \vee \bar{\mathcal{E}}(r, \phi) \wedge \bar{\mathcal{C}}(\phi),
        \end{equation}
        \noindent with $\bar{\mathcal{D}}(r, \phi) = {\mathcal{D}}_3(r, \phi) \vee {\mathcal{D}}_4(r, \phi)$ and $\bar{\mathcal{E}}(r, \phi) = {\mathcal{E}}_3(r, \phi) \vee {\mathcal{E}}_4(r, \phi)$. Then, it can be shown that $\bar{\mathcal{B}}(r,\phi)$ can be expressed as
        \begin{equation}
            \bar{\mathcal{B}}(r,\phi) = \underbrace{\mathcal{D}_3(r,\phi) \wedge \mathcal{C}_1(\phi)}_{\mathcal{M}_1(r,\phi)} \vee ... \vee 
            \underbrace{\mathcal{E}_4(r,\phi) \wedge \mathcal{C}_4(\phi)}_{\mathcal{M}_8(r,\phi)},
        \end{equation}
        \noindent being $\mathfrak{M}_i(r) = \{ \phi \in [0,2\pi] \lvert \mathcal{M}_i(r, \phi) = \mathtt{1}$, where the details are omitted due to space limitations. 
        
\section{Proof of Corollary \ref{cor:Disjoint_set}}
\label{App:Proof of cor:Disjoint_set}
        The proof comes after realizing that the boolean expression $\mathcal{C}_1(\phi)$ and $\mathcal{C}_2(\phi)$, which are given in \eqref{eq:C1 and C2}, are restricted to the events $\mathcal{A}_t(\phi)$ and $\bar{\mathcal{A}_t(\phi)}$, respectively; the expressions $\mathcal{D}_1(\phi)$ and $\mathcal{D}_2(\phi)$ from \eqref{eq:D1 and D2} to the events $\mathcal{A}_c(\phi)$ and $\bar{\mathcal{A}_c(\phi)}$; $\mathcal{E}_1(\phi)$ and $\mathcal{E}_2(\phi)$ from \eqref{eq:E1 and E2} to the events $\mathcal{A}_s(\phi)$ and $\bar{\mathcal{A}_s(\phi)}$; and finally $\mathcal{C}_3(\phi)$ and $\mathcal{C}_4(\phi)$ 
        to $\mathcal{A}_t(\phi)$ and $\bar{\mathcal{A}_t(\phi)}$. 
        Expressing these events as intervals of the angle $\phi$ and substituting on each of the boolean expressions $\mathcal{X}_i(r,\phi)$ in \eqref{eq:final_proof_B} completes the proof. The same process is followed for the sets $\mathfrak{M}_i(r)$ with $i \in [1,8] \subset \mathbb{Z}$.

\section{Proof of Theorem \ref{theor:Joint CDF}}
\label{App:Proof of theor:Joint CDF}
    Using \textbf{Lemma \ref{lem:Rectangle_in_polar}} and \ref{lem:Step2}, the overlap area given by \eqref{eq:int_convex} can be written as follows
    \begin{align}
        \label{eq:overlap_Z}
        \lvert \mathfrak{Z}(u,r,\theta) \lvert &= \frac{r^2}{2} \int_{\phi=0}^{\theta} \mathbbm{1}_{\mathfrak{T}_{\mathfrak{B}(r)}} (\phi) \dd{\phi} \\
        &+ \int_{\phi=0}^{\theta} \frac{h_x^2(\phi)}{\cos^2(\phi)} \mathbbm{1}(\mathcal{C}(\phi) \wedge \bar{\mathcal{B}}(r,\phi)) \dd{\phi} \\ \nonumber
        &+ \int_{\phi=0}^{\theta} \frac{h_y^2(\phi)}{\sin^2(\phi)} \mathbbm{1}(\bar{\mathcal{C}}(\phi) \wedge \bar{\mathcal{B}}(r,\phi)) \dd{\phi}, \nonumber
    \end{align} 
    \noindent where the equality $\mathbbm{1}({\mathcal{B}}(r,\phi)) = \mathbbm{1}_{\mathfrak{T}_{\mathfrak{B}(r)}} (\phi)$ is used, and the term $\beta(\phi)$ in \eqref{eq:int_convex} is expressed as the sum of two indicator functions.
    The term $\mathbbm{1}(\mathcal{C}(\phi) \wedge \bar{\mathcal{B}}(r,\phi))$ can be manipulated as follows
    \begin{align}
        \label{eq:M_1_4}
        \mathbbm{1}(\mathcal{C}(\phi) &\wedge \bar{\mathcal{B}}(r,\phi))   \\
            & \overset{\mathrm{(a)}}{=} \mathbbm{1}(\mathcal{C}(\phi) \wedge (\bar{\mathcal{D}}(r,\phi) \wedge \mathcal{C}(\phi) \vee \bar{\mathcal{E}}(r,\phi) \wedge \bar{\mathcal{C}}(\phi)))
            \nonumber \\
            & \overset{\mathrm{(b)}}{=} \mathbbm{1}( (\mathcal{D}_3(r,\phi) \vee \mathcal{D}_4(r,\phi)) \wedge (\mathcal{C}_1(\phi) \vee \mathcal{C}_2(\phi)))
            \nonumber \\ 
            & \overset{\mathrm{(c)}}{=}  \mathbbm{1}\left( \bigvee_{i_1}^{4} \mathcal{M}_i(r,\phi) \right) = \sum_{i=1}^{4} \mathbbm{1}\left(  \mathfrak{M}_i(r,\phi) \right), \nonumber
    \end{align} 
    \noindent where (a) comes after expressing $\bar{\mathcal{B}}(r,\phi))$ as $\bar{\mathcal{D}}(r,\phi) \wedge \mathcal{C}(\phi) \vee \bar{\mathcal{E}}(r,\phi) \wedge \bar{\mathcal{C}}(\phi)$; (b) after applying distributive, associative and absorption properties of the logical operators; and (c) after some manipulations, applying \textbf{Corollary \ref{cor:Disjoint_set}} and identifying the terms $\mathcal{D}_k(r,\phi) \wedge \mathcal{C}_j(\phi) = \mathcal{M}_\ell$ with $k\in\{3,4\}$, $j\in\{1,2\}$ and $\ell\in\{1,2,3,4\}$. 
    
    Analogously, the term $\mathbbm{1}(\bar{\mathcal{C}}(\phi) \wedge \bar{\mathcal{B}}(r,\phi))$ can be written as
    \begin{align}
        \label{eq:M_5_8}
        \mathbbm{1}(\bar{\mathcal{C}}(\phi) \wedge \bar{\mathcal{B}}(r,\phi)) = \sum_{i=5}^{8} \mathbbm{1}\left(  \mathfrak{M}_i(r,\phi) \right).
    \end{align}     
    Finally, substituting \eqref{eq:M_5_8} and \eqref{eq:M_1_4} in \eqref{eq:overlap_Z} and \eqref{eq:F_R_Theta} and applying \textbf{Lemma \ref{lem:Step2}}, \textbf{Proposition \ref{prop:Definite_int}} and \textbf{Corollary \ref{cor:Definite_int}} completes the proof after some additional manipulations. 
    
\section{Proof of Corollaries \ref{cor:joint_pdf_R_Theta} and \ref{cor:marginal_azimuth_CDF}}
\label{App:Proof_corollaries}   

\subsection{Corollary \ref{cor:joint_pdf_R_Theta}}
    The joint PDF of distance and azimuth angle is computed as $f_{R,\Theta} (r, \theta) = \pdv{F_{R,\Theta} (r, \theta)}{r}{\theta}$. It can be noticed that the derivative of the two summations multiplied by $\frac{h_i^2}{2 \ell_x \ell_y}$ in \eqref{eq:Joint_CDF} are $0$ since they do not have terms that depend simultaneously on the $r$ and $\theta$ variables. The partial derivative with respect to $\theta$ of the terms $\Big(\min(\theta, \chi_{i,2}(r)) - x_{i,1}(r)\Big)^+$ can be written as follows
        \begin{align}
            \label{eq:intermediate_result_joint_PDF}
            \pdv{}{\theta} & \Big(\min(\theta, \chi_{i,2}(r)) -  \chi_{i,1}(r)\Big)^+ =  \\
            & \mathbbm{1} \left( \theta < \chi_{i,2}(r)) \right) \mathbbm{1} \left(\min(\theta, \chi_{i,2}(r)) > \chi_{i,1}(r) \right), \nonumber
        \end{align}
        \noindent where the term $\min(\theta, \chi_{i,2}(r))$ has been expressed as $\theta \mathbbm{1} \left( \theta < \chi_{i,2}(r) \right)+  \chi_{i,2}(r) \mathbbm{1} \left( \theta \geq \chi_{i,2}(r)) \right)$, it has been computed the derivative with respect to $\theta$ and it has been multiplied by $\mathbbm{1}\left(\min(\theta, \chi_{i,2}(r)) > \chi_{i,1}(r) \right)$ as per \textbf{Definition \ref{def:+ operator}}. The superscript $(<)$ or $(\geq)$ in $\chi_{i,1}(r)$ and $\chi_{i,2}(r)$ has been omitted to refer to both cases. 
        Finally, multiplying by $\frac{r^2}{2 \ell_x \ell_y}$, deriving with respect to $r$ and manipulating the resulting expression completes the proof.
        
\subsection{Corollary \ref{cor:marginal_azimuth_CDF}}
    The marginal CDF is computed as $F_{\Theta}(\theta) = \lim\limits_{r\to\infty} F_{R, \Theta}(r, \theta)$. It can be shown that the summation multiplied by $\frac{r^2}{2 \ell_x \ell_y}$ in \eqref{eq:Joint_CDF} is $0$ since the terms multiplied by $\mathbbm{1} (r < h_i)$ are $0$ due to the indicator function, and 
    the terms multiplied by $\mathbbm{1} (r \geq h_i)$ are $0$ since the argument of the $(\bullet)^+$ operator are negative when $r \to \infty$. 
    This can be checked from \eqref{eq:chi_i(r)} since \mbox{ $\lim\limits_{r\to\infty} \acos(\frac{h_i}{r}) = \frac{\pi}{2}$} and $\lim\limits_{r\to\infty} \asin(\frac{h_i}{r}) = 0$ and $\atan(x) \in [0,\frac{\pi}{2})$ when $x\geq 0$ whereas $\atan(x) \in (-\frac{\pi}{2}, 0)$ when $x < 0$.
    Finally, deriving the limit when $r \to \infty$ of the term $\mu_{i,1}(r)$ and $\min(\theta, \mu_{i,2}(r))$ on the two summations multiplied by $\frac{h_i^2}{2 \ell_x \ell_y}$ completes the proof.
    
\section{Proof of Theorem \ref{theo:Joint_CDF_3D}}
\label{App:Proof of theo:Joint_CDF_3D}

        
        
        The determination of the distribution of distance, azimuth and zenith angles can be posed as a standard RV transformation problem from the polar coordinates, $R$, and $\Theta$ of the 2D case. With this approach, the RV transformation is written as follows:
        \begin{align}
            \label{eq:3D_transformation}
            D &= f(R, u_z, v_z) = \sqrt{R^2 + (u_z - v_z)^2}, \nonumber \\
            \Theta &= \Theta, \\
            \Psi &= g(R, u_z, v_z) = 
            \begin{cases}
                \pi - \atan\left(\frac{R}{u_z-v_z} \right)    & \mathrm{if} \; u_z \geq v_z \\
                \atan\left(\frac{R}{v_z-v_z} \right)    & \mathrm{if} \; u_z < v_z.
            \end{cases} \nonumber 
        \end{align}
        Hence, the CDF of the distance and angle distribution can be expressed as follows
        \begin{align}
            \label{eq:3D_case_integral}
            & F_{D, \Theta, \Psi} (\mathscr{d}, \theta, \psi) = 
            \nonumber \\ 
            & \quad \Pr \Big(\overbrace{f(R, u_z, v_z) \leq \mathscr{d}}^{\mathcal{F}(\mathscr{d})}, \Theta \leq \theta, \overbrace{g(R, u_z, v_z) \leq \psi}^{\mathcal{G}(\psi)}\Big)
            \nonumber \\ 
            & \quad = \int_{\theta^\prime = 0}^{\theta} \int_{r>0} \mathbbm{1} \left(\mathcal{F}(\mathscr{d}) \right) \mathbbm{1} \left(\mathcal{G}(\psi) \right) f_{R,\Theta} (r, \theta^\prime) \dd{r} \dd{\theta^\prime}.
        \end{align}
        The next step is to isolate the variable $R$ on the Boolean expressions $\mathcal{F}(\mathscr{d})$ and $\mathcal{G}(\psi)$. 
        %
        %
        Finally, 
         applying the indicator functions over the limits of the integrals, and identifying the resulting expression in terms the joint CDF of distance and azimuth angle completes the proof. 


\bibliographystyle{IEEEtran}
\bibliography{references.bib}

\begin{thebibliography}{10}
\providecommand{\url}[1]{#1}
\csname url@samestyle\endcsname
\providecommand{\newblock}{\relax}
\providecommand{\bibinfo}[2]{#2}
\providecommand{\BIBentrySTDinterwordspacing}{\spaceskip=0pt\relax}
\providecommand{\BIBentryALTinterwordstretchfactor}{4}
\providecommand{\BIBentryALTinterwordspacing}{\spaceskip=\fontdimen2\font plus
\BIBentryALTinterwordstretchfactor\fontdimen3\font minus
  \fontdimen4\font\relax}
\providecommand{\BIBforeignlanguage}[2]{{%
\expandafter\ifx\csname l@#1\endcsname\relax
\typeout{** WARNING: IEEEtran.bst: No hyphenation pattern has been}%
\typeout{** loaded for the language `#1'. Using the pattern for}%
\typeout{** the default language instead.}%
\else
\language=\csname l@#1\endcsname
\fi
#2}}
\providecommand{\BIBdecl}{\relax}
\BIBdecl

\bibitem{ElSawy17}
H.~ElSawy and et~al., ``{Modeling and Analysis of Cellular Networks Using
  Stochastic Geometry: A Tutorial},'' \emph{IEEE Commun. Surveys Tuts.},
  vol.~19, no.~1, pp. 167--203, 2017.

\bibitem{DiRenzo21}
Y.~Hmamouche and et~al., ``{New Trends in Stochastic Geometry for Wireless
  Networks: A Tutorial and Survey},'' \emph{Proc. IEEE}, vol. 109, 2021.

\bibitem{Martin18_Mag}
F.~J. Martin-Vega and et~al., ``{Key Technologies, Modeling Approaches, and
  Challenges for Millimeter-Wave Vehicular Communications},'' \emph{IEEE
  Commun. Mag.}, vol.~56, no.~10, pp. 28--35, 2018.

\bibitem{Shi21}
M.~Shi and et~al., ``{Meta Distribution of the SINR for mmWave Cellular
  Networks With Clusters},'' \emph{IEEE Trans. Commun.}, vol.~69, no.~10, 2021.

\bibitem{Kalamkar21}
S.~S. Kalamkar and et~al., ``{Beam Management in 5G: A Stochastic Geometry
  Analysis},'' \emph{IEEE Trans. Wireless Commun.}, pp. 1--1, 2021.

\bibitem{Rebato19}
M.~Rebato and et~al., ``{Stochastic Geometric Coverage Analysis in mmWave
  Cellular Networks With Realistic Channel and Antenna Radiation Models},''
  \emph{IEEE Trans. Commun.}, vol.~67, no.~5, 2019.

\bibitem{Singh21}
G.~Singh and et~al., ``{Stochastic Geometry-Based Interference Characterization
  for RF and VLC-Based Vehicular Communication System},'' \emph{IEEE Syst. J.},
  vol.~15, no.~2, 2021.

\bibitem{Ammar20}
H.~A. Ammar and et~al., ``{A Poisson Line Process-Based Framework for
  Determining the Needed RSU Density and Relaying Hops in Vehicular
  Networks},'' \emph{IEEE Trans. Wireless Commun.}, vol.~19, no.~10, 2020.

\bibitem{Yuanwei19}
W.~Yi and et~al., ``{Modeling and Analysis of MmWave V2X Networks With
  Vehicular Platoon Systems},'' \emph{IEEE J. Sel. Areas Commun.}, vol.~37,
  no.~12, 2019.

\bibitem{Kusaladharma19}
S.~Kusaladharma, Z.~Zhang, and C.~Tellambura, ``{Interference and Outage
  Analysis of Random D2D Networks Underlaying Millimeter-Wave Cellular
  Networks},'' \emph{IEEE Trans. Commun.}, vol.~67, no.~1, pp. 778--790, 2019.

\bibitem{Haenggi18}
N.~Deng, M.~Haenggi, and Y.~Sun, ``{Millimeter-Wave Device-to-Device Networks
  With Heterogeneous Antenna Arrays},'' \emph{IEEE Trans. Commun.}, vol.~66,
  no.~9, pp. 4271--4285, 2018.

\bibitem{Shuanshuan18}
S.~Wu, R.~Atat, N.~Mastronarde, and L.~Liu, ``{Improving the Coverage and
  Spectral Efficiency of Millimeter-Wave Cellular Networks Using
  Device-to-Device Relays},'' \emph{IEEE Trans. Commun.}, vol.~66, no.~5, pp.
  2251--2265, 2018.

\bibitem{Maeng21}
S.~J. Maeng, M.~A. Deshmukh, I.~Güvenç, A.~Bhuyan, and H.~Dai,
  ``{Interference Analysis and Mitigation for Aerial IoT Considering 3D Antenna
  Patterns},'' \emph{IEEE Trans. Veh. Technol.}, vol.~70, no.~1, pp. 490--503,
  2021.

\bibitem{Azari20}
M.~M. Azari, G.~Geraci, A.~Garcia-Rodriguez, and S.~Pollin, ``{UAV-to-UAV
  Communications in Cellular Networks},'' \emph{IEEE Trans. Wireless Commun.},
  vol.~19, no.~9, pp. 6130--6144, 2020.

\bibitem{Enayati19}
S.~Enayati, H.~Saeedi, H.~Pishro-Nik, and H.~Yanikomeroglu, ``{Moving Aerial
  Base Station Networks: A Stochastic Geometry Analysis and Design
  Perspective},'' \emph{IEEE Trans. Wireless Commun.}, vol.~18, no.~6, pp.
  2977--2988, 2019.

\bibitem{Lyu21}
J.~Lyu and R.~Zhang, ``{Hybrid Active/Passive Wireless Network Aided by
  Intelligent Reflecting Surface: System Modeling and Performance Analysis},''
  \emph{IEEE Trans. Wireless Commun.}, pp. 1--1, 2021.

\bibitem{Alouini21}
M.~A. Kishk and M.-S. Alouini, ``{Exploiting Randomly Located Blockages for
  Large-Scale Deployment of Intelligent Surfaces},'' \emph{IEEE J. Sel. Areas
  Commun.}, vol.~39, no.~4, pp. 1043--1056, 2021.

\bibitem{Nemati21}
M.~Nemati, B.~Maham, S.~R. Pokhrel, and J.~Choi, ``{Modeling RIS Empowered
  Outdoor-to-Indoor Communication in mmWave Cellular Networks},'' \emph{IEEE
  Trans. Commun.}, pp. 1--1, 2021.

\bibitem{Afshang17}
M.~Afshang and H.~S. Dhillon, ``{Fundamentals of Modeling Finite Wireless
  Networks Using Binomial Point Process},'' \emph{IEEE Trans. Wireless
  Commun.}, vol.~16, no.~5, pp. 3355--3370, 2017.

\bibitem{DaSilva20}
J.~Kibiłda, A.~B. MacKenzie, M.~J. Abdel-Rahman, S.~K. Yoo, L.~G. Giordano,
  S.~L. Cotton, N.~Marchetti, W.~Saad, W.~G. Scanlon, A.~Garcia-Rodriguez,
  D.~López-Pérez, H.~Claussen, and L.~A. DaSilva, ``{Indoor Millimeter-Wave
  Systems: Design and Performance Evaluation},'' \emph{Proc. IEEE}, vol. 108,
  no.~6, pp. 923--944, 2020.

\bibitem{Kokkoniemi20}
J.~Kokkoniemi, A.-A.~A. Boulogeorgos, M.~U. Aminu, J.~Lehtomäki, A.~Alexiou,
  and M.~Juntti, ``{Stochastic Analysis of Indoor THz Uplink with Co-Channel
  Interference and Phase Noise},'' in \emph{2020 IEEE International Conference
  on Communications Workshops (ICC Workshops)}, 2020, pp. 1--6.

\bibitem{Abarghouyi18a}
S.~M. Azimi-Abarghouyi, B.~Makki, M.~Haenggi, M.~Nasiri-Kenari, and
  T.~Svensson, ``Stochastic geometry modeling and analysis of single- and
  multi-cluster wireless networks,'' \emph{IEEE Transactions on
  Communications}, vol.~66, no.~10, pp. 4981--4996, 2018.

\bibitem{Azimi-Abarghouyi19}
S.~M. Azimi-Abarghouyi, B.~Makki, M.~Nasiri-Kenari, and T.~Svensson,
  ``{Stochastic Geometry Modeling and Analysis of Finite Millimeter Wave
  Wireless Networks},'' \emph{IEEE Trans. Veh. Technol.}, vol.~68, no.~2, pp.
  1378--1393, 2019.

\bibitem{Srinivasa10}
S.~Srinivasa and M.~Haenggi, ``{Distance Distributions in Finite Uniformly
  Random Networks: Theory and Applications},'' \emph{IEEE Trans. Veh.
  Technol.}, vol.~59, no.~2, pp. 940--949, 2010.

\bibitem{Khalid13}
Z.~Khalid and S.~Durrani, ``{Distance Distributions in Regular Polygons},''
  \emph{IEEE Trans. Veh. Technol.}, vol.~62, no.~5, pp. 2363--2368, 2013.

\bibitem{Fan07}
P.~Fan, G.~Li, K.~Cai, and K.~B. Letaief, ``{On the Geometrical Characteristic
  of Wireless Ad-Hoc Networks and its Application in Network Performance
  Analysis},'' \emph{IEEE Trans. Wireless Commun.}, vol.~6, no.~4, pp.
  1256--1265, 2007.

\bibitem{Haenggi12}
M.~Haenggi, \emph{Stochastic Geometry for Wireless Networks}, 1st~ed.\hskip 1em
  plus 0.5em minus 0.4em\relax USA: Cambridge University Press, 2012.

\bibitem{Boyd06}
\BIBentryALTinterwordspacing
S.~Boyd and L.~Vandenberghe, \emph{Convex Optimization}.\hskip 1em plus 0.5em
  minus 0.4em\relax {Cambridge University Press}, March 2004. [Online].
  Available:
  \url{http://www.amazon.com/exec/obidos/redirect?tag=citeulike-20\&path=ASIN/0521833787}
\BIBentrySTDinterwordspacing

\bibitem{Banagar21}
M.~Banagar and H.~S. Dhillon, ``3d two-hop cellular networks with wireless
  backhauled uavs: Modeling and fundamentals,'' \emph{IEEE Trans. Wireless
  Commun.}, pp. 1--1, 2022.

\bibitem{Balanis12}
C.~A. Balanis, \emph{Antenna theory: analysis and design}.\hskip 1em plus 0.5em
  minus 0.4em\relax Wiley-Interscience, 2005.

\bibitem{Bjornson17}
\BIBentryALTinterwordspacing
E.~Björnson, J.~Hoydis, and L.~Sanguinetti, ``Massive mimo networks: Spectral,
  energy, and hardware efficiency,'' \emph{Foundations and Trends® in Signal
  Processing}, vol.~11, no. 3-4, pp. 154--655, 2017. [Online]. Available:
  \url{http://dx.doi.org/10.1561/2000000093}
\BIBentrySTDinterwordspacing

\bibitem{Giordani19}
M.~Giordani, M.~Polese, A.~Roy, D.~Castor, and M.~Zorzi, ``A tutorial on beam
  management for 3gpp nr at mmwave frequencies,'' \emph{IEEE Communications
  Surveys Tutorials}, vol.~21, no.~1, pp. 173--196, 2019.

\bibitem{38.901}
3GPP, \emph{{Technical Report (TR); Study on channel model for frequencies from
  0.5 to 100 GHz}}, {3rd Generation Partnership Project (3GPP)} TR {38.901},
  Rev. 17.0.0, April 2022.

\end{thebibliography}

\end{document}